\documentclass[useAMS,usenatbib,a4paper]{mn2e}

\usepackage[dvips]{graphicx}
\usepackage{amssymb}
\usepackage{txfonts}
\usepackage{color}

\newcommand{\thisgrb}{GRB\,100901A}

\newcommand{\zd}{[O\,{\sc iii}]}
\newcommand{\zt}{[O\,{\sc ii}]}

\title[The afterglow spectrum of \thisgrb]{The host-galaxy response to the afterglow of \thisgrb\thanks{Based on observations obtained at the Gemini Observatory at Mauna Kea under programme GN-2010B-Q-7, and observations obtained with ESO Telescopes at the Paranal Observatory under programme 085.A-0009(B).}}

\author[Olga E. Hartoog et al.]
{\parbox{\textwidth}{Olga E.~Hartoog$^{1,2}$ \thanks{O.E.Hartoog@uva.nl}, 
Klaas~Wiersema$^{3}$,
Paul~M.~Vreeswijk$^{4}$, 
Lex~Kaper$^{1}$, 
Nial~R.~Tanvir$^{3}$,
\mbox{Sandra}~Savaglio$^{5}$, 
Edo~Berger$^{6}$, 
Ryan~Chornock$^{5}$, 
\mbox{Stefano}~Covino$^{7}$, 
Valerio~D'Elia$^{8,9}$, 
Hector~Flores$^{10}$, 
Johan~P.~U.~Fynbo$^{11}$,
Paolo~Goldoni$^{12}$,
Andreja~Gomboc$^{13,14}$,
Andrea~Melandri$^{7,15}$, 
Alexei~Pozanenko$^{16}$, 
Joop~Schaye$^{2}$,
Antonio~de~Ugarte~Postigo$^{17}$, 
Ralph~A.~M.~J.~Wijers$^{1}$\\
}\vspace{0.3cm}\\
\parbox{\textwidth}{(Affiliations can be found after the references)
}}

\begin{document}

\pagerange{\pageref{firstpage}--\pageref{lastpage}} \pubyear{2013}

\maketitle

\label{firstpage}

\begin{abstract}
\par For Gamma-Ray Burst 100901A, we have obtained Gemini-North and Very Large Telescope optical afterglow spectra at four epochs: one hour, one day, three days and one week after the burst, thanks to the afterglow remaining unusually bright at late times. Apart from a wealth of metal resonance lines, we also detect lines arising from fine-structure levels of the ground state of Fe\,{\sc ii}, and from metastable levels of Fe\,{\sc ii} and Ni\,{\sc ii} at the host redshift ($z=1.4084$). These lines are found to vary significantly in time. The combination of the data and modelling results shows that we detect the fall of the Ni\,{\sc ii}\,$^4F_{9/2}$ metastable level population, which to date has not been observed. Assuming that the population of the excited states is due to the UV-radiation of the afterglow, we estimate an absorber distance of a few hundred pc. This appears to be a typical value when compared to similar studies. We detect two intervening absorbers ($z=1.3147,\,1.3179$). Despite the wide temporal range of the data, we do not see significant variation in the absorption lines of these two intervening systems. 
\end{abstract}

\begin{keywords}
gamma-ray burst: individual: GRB\,100901A; gamma-rays: bursts - galaxies: abundances - galaxies: ISM - galaxies: distances and redshifts
\end{keywords}

\section{Introduction \label{sec:intro}}

\par Shortly after the first detection of an optical afterglow associated with a long gamma-ray burst \citep[GRB,][]{paradijs1997}, it became clear that GRBs may be useful probes of distant galaxies: afterglows can be very bright and have a simple power law continuum at ultraviolet (UV) to optical wavelengths, against which otherwise undetectable absorbing systems, such as the gas in the host galaxy, leave an observable signature \citep[see e.g.,][]{vreeswijk2001,savaglio2003,vreeswijk2004,berger2006,fynbo2006,prochaska2007}. Long GRBs are formed as end products of the evolution of massive stars \citep{woosley1993,paczynski1998} and should therefore probe star-forming regions. Their optical afterglows fade away on short timescales ($\sim$days), making it possible to study their host galaxies in emission as well \citep[see e.g.,][]{savaglio2009,hjorth2012}. For a review of long GRBs and their host galaxies, see e.g., \citet{savaglio2006,savaglio2012}.\\

\par Optical spectra (i.e., rest frame UV at redshifts around $z\sim1-2$) of GRB afterglows generally show strong absorption lines of metals, both at low and high-ionisation stages, as well as high neutral hydrogen column densities (see \citealt{fynbo2009} for a large and relatively unbiased sample). Absorption lines provide the redshift and allow estimation of column densities of the host galaxy of the GRB and possibly other absorbers along the line of sight. This provides valuable information on the metallicity and dust content of galaxies at arbitrary redshift, regardless of the brightness of these systems, which is very difficult to obtain otherwise. The highest redshift lines in the afterglow spectrum formally only provide a lower limit on the host redshift. However, the presence of fine-structure lines can confirm the galaxy as the host, because these arise from excited states that are expected to have been populated by the UV-afterglow radiation of the GRB \citep{prochaska2006a,vreeswijk2007}. These lines are not seen in foreground line-of-sight absorbers in quasar (QSO) spectra, except for some relatively low-energy excited states from carbon. Due to the transient nature of the afterglow, the absorption lines from these excited states are expected to vary in strength, which may provide information on the internal distribution of gas and kinematics within the host galaxy; often the only way to obtain such information for these distant and mostly faint galaxies. Fine-structure line variation has been measured and modelled for a handful of bursts (e.g., \citealt{vreeswijk2007,dessauges2006,d'elia2009a,ledoux2009,deugartepostigo2011,decia2012,vreeswijk2013}), which allowed constraints to be placed on the distance between the burst and the absorbing material (see \citealt{vreeswijk2013} for a description of the methods used).
\par The primary difficulty in probing column density variations in GRB sightlines to fit to the models described above, is the requirement on afterglow brightness: for fainter afterglows the necessary high signal-to-noise and high spectral resolution are difficult to obtain with current instrumentation. This also limits this method to early-time data and a limited range of instruments, making it difficult to build up a sample. In the cases described above, the variation is measured with high resolution spectrographs, and all spectra are obtained within a few hours post burst (observer's frame). 
\par The optical afterglow of \thisgrb\,remained unusually bright out to very late times, which allowed us to collect spectra over a time span of a week with the Gemini-North Multi-Object Spectrograph and X-shooter on the ESO Very Large Telescope. In this paper we investigate the behaviour of the excited level populations over this unusually long time span, and the application of the excitation models to low and intermediate resolution spectra. The observations and data reduction are described in Section~\ref{sec:obs}. We study the metal resonance lines at the host-galaxy redshift in Section~\ref{sec:resonancelines}. In Section~\ref{sec:hostgalaxy} we examine the dust content of the host galaxy. The two intervening absorbers are analysed in Section~\ref{sec:interv}. In Section~\ref{sec:varyfinestruc} the fine-structure line variability is described and modelled. In Section~\ref{sec:emission} we briefly discuss the emission lines from the host galaxy. We discuss our results in Section~\ref{sec:discus} and summarise our conclusions in Section~\ref{sec:conclusions}.
\par Throughout the paper we adopt a standard $\Lambda$CDM cosmology with $H_0 =  71$ km\,s$^{-1}$\,Mpc$^{-1}$,  $\Omega_\mathrm{m} = 0.27$, $\Omega_\Lambda = 0.73$. We indicate line transitions by their vacuum wavelengths.

\begin{table*}
\caption{Overview of the spectroscopic observations of the afterglow of \thisgrb, arranged by instrument and epoch. Columns (1) and (2) indicate telescope, instrument, setup, and slit dimensions. Column (3) gives the wavelength coverage for this setup and the resolving power $R=\lambda/\Delta\lambda$, with $\Delta\lambda$ the FWHM of an unresolved line. Column (4) gives the mid time of the epochs of observation in days after the burst BAT trigger. Column (5) lists the interpolated observed $R$-magnitude of the afterglow at this epoch (Gomboc et al., in prep). Column (6) lists the total exposure time of the observation in this epoch. Columns (7) and (8) give the seeing and the airmass. The seeing is measured by fitting a gaussian along the spatial direction of the slit, around the central wavelength of each spectrum. Column (9) gives the signal-to-noise ratio (S/N) in the continuum at 4150 and 6100\,\AA\,per GMOS pixel ($\delta \lambda=0.91$\,\AA). Column (10) gives the name we will use to refer to the epoch.
\label{tab:logobs}}
\begin{tabular}{llllllllll} 
Telescope		& Instrument  			&Spectral Range 		&Time after Burst 	& $m_R$	&Exp. Time & Seeing		& Airmass & S/N$/(\delta \lambda=0.91\AA)$ 			& Abbrev.\\ 
			& and Setup			& and Resolution		& (days)			& 	 	&(s)		& ($^{\prime\prime}$)			&		& at $4150,\,6100\,\AA$	&	\\
\hline
Gemini-North 	& GMOS, 				&$3810-6710$ \AA			& 0.0526		&17.8 &$4 \times 400$	&  0.62		& 1.06	& 38, 95		& epoch1  \\
                          & B600+\,G5307  	    	&$R=1000-1900$ 	& 1.0095					& 18.6&$4 \times 500$	&  0.52		& 1.01 	& 16, 70			& epoch2 \\ 
		        & $0.75^{\prime\prime} \times 330^{\prime\prime}$		&	& 7.0270		& 22.3&$4 \times 1200$	&  0.68		& 1.05  	& 5, 13			& epoch3 \\ 
Very Large	& X-Shooter 	&$3000-25000$ \AA 		& 2.7492					& 20.2&$4 \times 600$	&  1.30		& 1.48 	& 9, 7			& xsh	 \\
Telescope		&UVB $1.0^{\prime\prime} \times 11^{\prime\prime}$ 	&$R=5400$			&			& &				&			&		&			&\\
			&VIS $0.9^{\prime\prime} \times 11^{\prime\prime}$	&$R=7400$			&			&&				&			&		&			&\\
			&NIR $0.9^{\prime\prime} \times 11^{\prime\prime}$ 	&$R=5800$			&			&&				&			&		&			&\\

\hline
\end{tabular}
\end{table*}

\section{Observations}\label{sec:obs}
\thisgrb\,was detected by \emph{Swift} on 2010 September 1 at 13:34:10 UT. With a total duration of $T_{90}=439 \pm 33$\,s \citep{immler2010a,sakamoto2010} the burst is clearly classified as a long burst. The optical afterglow candidate at RA (J2000) = 01$^\mathrm{h}$49$^\mathrm{m}$03.42$^\mathrm{s}$, Dec =+22$^{\circ}$45$^{\prime}$30.8$^{\prime\prime}$ (90\% confidence error radius of about $0.81^{\prime\prime}$) was identified in an Ultraviolet/Optical Telescope (UVOT) exposure which started 147\,s after the Burst Alert Telescope (BAT) trigger \citep{immler2010a}. Automatic observations with the Faulkes Telescope North identified an uncatalogued object at the position consistent with the UVOT candidate \citep{guidorzi2010}. Several follow-up photometry efforts followed, resulting in the light curve presented and analysed in Gomboc et al. (in prep) and \citet{gorbovskoy2012} (see also Figure~\ref{fig:finestruc_variability}). The first optical spectrum of the afterglow was taken with the Gemini Multi-Object Spectrograph (GMOS) on Gemini-North (programme GN-2010B-Q-7, PI Tanvir), approximately 1\,hr 15 mins (0.0526 days) after burst trigger \citep{chornock2010}. With this instrument, two other spectra at respectively 1 and 7 days after the burst were obtained. A fourth spectrum was obtained approximately 3 days after the burst with the X-shooter spectrograph mounted on the ESO Very Large Telescope (VLT), under programme 085.A-0009(B) (PI Fynbo). See Table~\ref{tab:logobs} for a detailed log of the observations presented in this paper.

	\subsection{Gemini-N/GMOS spectroscopy}
Gemini-N/GMOS is a low-resolution long slit spectrograph, equipped with three detectors. For the observations presented here, a slit width of $0.75^{\prime\prime}$ and the B600 grism with the G5307 order suppression filter have been used, which resulted in the wavelength coverage and resolving power reported in Table \ref{tab:logobs}. We used four exposures per epoch, using dithers in both dispersion and spatial coordinates to sample over the chip gaps and regions affected by amplifier location. The GMOS spectra have been reduced with the IRAF packages for Gemini GMOS (version 1.9), using arc line exposures taken directly before and after the science data. The four exposures are combined after extraction. The resulting spectra were normalised and no flux calibration was performed.

	\subsection{VLT/X-shooter spectroscopy}
	VLT/X-shooter is a cross-dispersed \'echelle spectrograph where the incoming light is split into three wavelength arms using dichroics, covering the full optical to near infrared wavelength range simultaneously \citep{d'odorico2006,vernet2011}. The X-shooter spectra have been taken in nodding mode with $1\times2$ binning (i.e., binning in the dispersion direction) in the UV-Blue (UVB) and Visual (VIS) arms, using a 100kHz / high-gain readout and $4\times600$s exposure times.
The selected slit widths were 1.0, 0.9 and $0.9^{\prime\prime}$ for the UVB, VIS and near-infrared (NIR) arms, respectively. A $5^{\prime\prime}$ nod throw along the slit was used to improve sky subtraction. The spectra were reduced using the ESO pipeline software version 1.2.2, using the so-called physical model mode  \citep{goldoni2006,modigliani2010}. We used calibration data (bias, dark, arc lamp, flat-field and flexure control frames) taken the next day or as close in time as possible to the science observations. The bin sizes of the resulting spectra were 0.2, 0.2, and 0.5\AA\,for UVB, VIS and NIR spectra, respectively. Extracted spectra were flux calibrated using spectrophotometric standard star exposures, also taken the same night, resulting in roughly flux-calibrated spectra spanning the near-ultraviolet to the near-infrared. We caution that the weather conditions during this night were poor, and likely non-photometric. We used a telluric standard star (HD\,4670, a B9\,V star) to correct the telluric absorption features in the NIR afterglow spectrum, using the IDL  {\sc spextool} package (\citealt{vacca2003}, see also \citealt{wiersema2011} for the use on X-shooter data).

\begin{figure*}
\includegraphics[width=18cm]{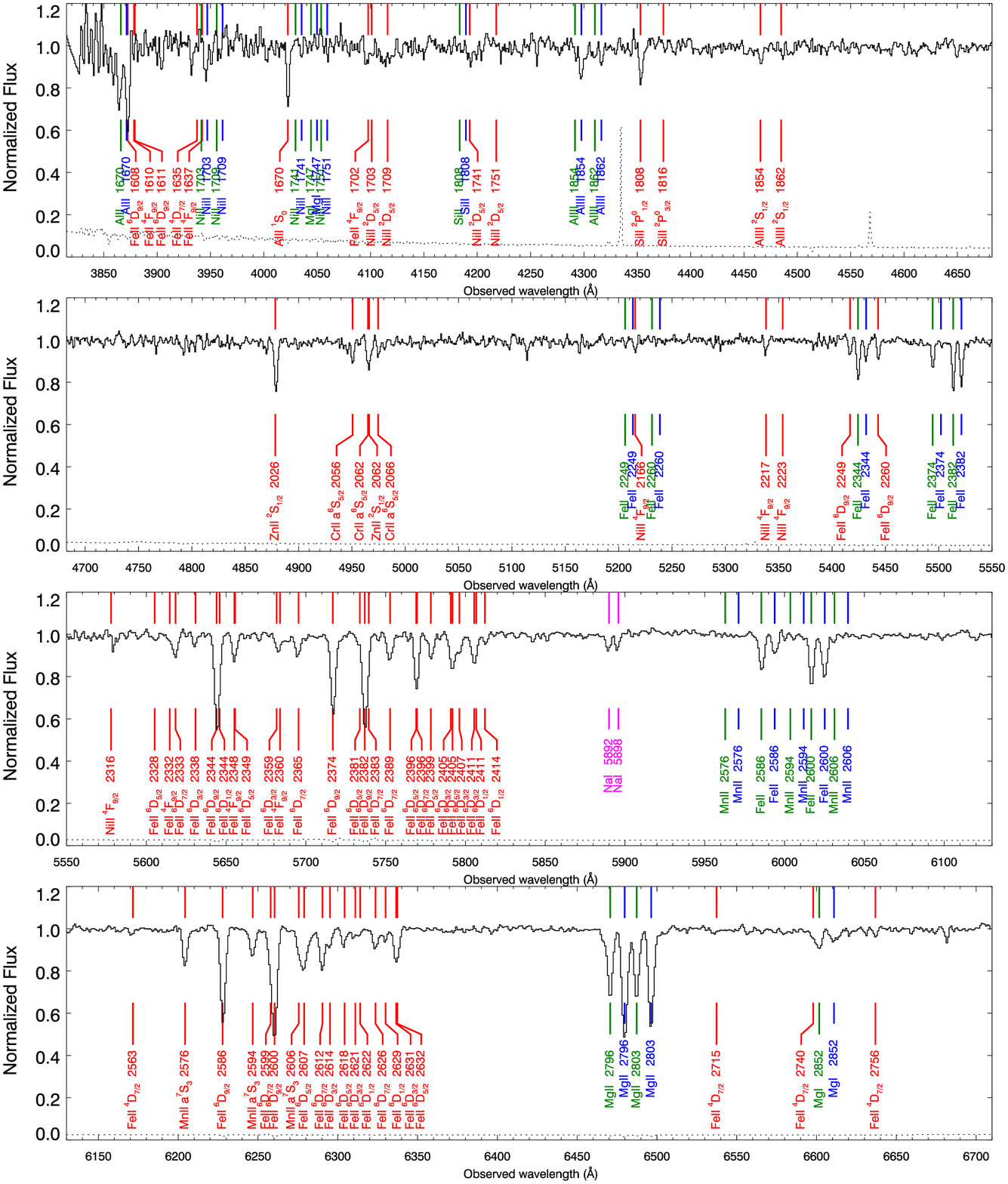} 
\caption{Normalised GMOS spectrum (epoch1) of the afterglow of \thisgrb. The red labels indicate the absorption lines of the host ($z_h=1.4084$); the green and blue labels (with vertical offset) indicate the absorption lines formed in two intervening systems at $z_1=1.3147$ and $z_2=1.3179$. The magenta labels show unshifted Na\,{\sc i} lines ($z=0$), resulting from the absorption in the Milky Way. The numbers at the labels indicate the configuration of the lower level and the vacuum rest wavelength of the transition.
The host shows resonance lines from the ions Mn\,{\sc ii}, Fe\,{\sc ii}, Cr\,{\sc ii}, Zn\,{\sc ii}, Si\,{\sc ii}, Al\,{\sc ii} and Al\,{\sc iii}, as well as lines from excited states of Fe\,{\sc ii} and Ni\,{\sc ii}, which are found to vary significantly (see Section~\ref{sec:varyfinestruc}).
In the intervening systems we detect Mg\,{\sc i}, Mg\,{\sc ii}, Fe\,{\sc ii}, Al\,{\sc ii} and Al\,{\sc iii}. 
\label{fig:epoch1}}
\end{figure*}

\section{Analysis}
The GMOS epoch1 spectrum (see Figure~\ref{fig:epoch1}) is that of a bright afterglow including many metal absorption lines. At a redshift of 1.4084 we detect strong resonance lines of Fe\,{\sc ii}, Mn\,{\sc ii}, Cr\,{\sc ii}, Al\,{\sc ii}, Al\,{\sc iii}, Zn\,{\sc ii} and Si\,{\sc ii}. Additionally, at the same redshift, line transitions arising from excited states of Fe\,{\sc ii} and Ni\,{\sc ii} are clearly detected (see Section~\ref{sec:varyfinestruc}). Therefore we identify this as the host-galaxy redshift $z_h$ \citep{chornock2010}. We do not detect lines from Fe\,{\sc iii}, as observed in the case of GRB\,080310 \citep{decia2012}. No Lyman-$\alpha$ is detected, because it is too far in the blue for the covered wavelength range. Two intervening absorbers are visible at $z_1=1.3147$ and $z_2=1.3179$ \citep{chornock2010}. In these systems we detect Fe\,{\sc ii}, Al\,{\sc ii}, Al\,{\sc iii}, Mg\,{\sc i} and Mg\,{\sc ii} (rest frame equivalent width $W_\lambda <1\,\AA$). We note that these lines are weaker at the intervening system redshifts than at the host redshift. There are no fine-structure lines detected at any of the intervening absorber redshifts. \mbox{Na\,{\sc i} $\lambda\lambda\, 5892,5898$} are detected at $z=0$ and are thus due to foreground absorption in the Galaxy.
\par The GMOS epoch2 spectrum (not fully shown in a figure) looks similar, though the signal-to-noise ratio (S/N) is lower due to the decreasing brightness of the afterglow. The line transitions arising from the metastable Ni\,{\sc ii} $^4F_{9/2}$ level (see Table~\ref{tab:Nitrans_metastab}) appear more pronounced in epoch2 than in epoch1; the Fe\,{\sc ii} fine-structure lines have become slightly weaker.
\par Since GMOS epoch3 is observed after the break in the light curve at $\sim2$ days (Gomboc et al., in prep), which causes the brightness to drop more quickly, the S/N is lower than for the other epochs, despite the longer integration time and comparable weather conditions. Still, strong metal resonance lines are clearly detected. Fine-structure lines that were present in epoch1 and lines from metastable levels that were present in epoch2 are much weaker or not detected any more (see Section~\ref{sec:varyfinestruc}).  
\par The quality of the X-shooter spectrum (xsh, after 2.75 days, see Table~\ref{tab:logobs}) is low due to poor weather conditions and large airmass, but because of the extraordinarily large wavelength coverage of this instrument, this spectrum does provide additional information. Moreover, the higher resolution allow us to exclude the existence of different velocity components down to $\sim70\,\mathrm{km\,s}^{-1}$ with respect to $\sim200\,\mathrm{km\,s}^{-1}$ for GMOS. The strong and saturated Mg\,{\sc ii}\,$\lambda\lambda\,2796,2803$, Mg\,{\sc i}\,$\lambda\,2853$, and Ca\,{\sc ii}\,$\lambda\lambda\,3934,3969$ absorption lines from the host galaxy are outside the range of GMOS due to the redshift, but they are clearly detected in xsh-VIS
 (not shown in a figure). Na\,{\sc i}\,$\lambda\,5892,5897$ falls in a region with atmospheric absorption and can therefore not be detected. We do not detect the red wing of the Ly-$\alpha$ absorption line due to low S/N. Furthermore, in xsh-NIR, we marginally detect forbidden oxygen emission line \zd$\,\lambda5007$ at the host-galaxy redshift (see Figure~\ref{fig:xshemission} and Section~\ref{sec:emission}).
\par We do not detect prominent interstellar (dust) extinction features such as the 2175\AA\,feature \citep[see e.g.,][]{eliasdottir2009} or diffuse interstellar bands in any of the spectra at the host-galaxy redshift. \\

	\subsection{Resonance lines}
	\label{sec:resonancelines}
	To measure the observed equivalent width $W_{\lambda,\mathrm{obs}}$ of the absorption lines in the spectrum, we locally fit gaussian functions to the normalised spectrum with {\sc ngaussfit}, a task in the {\sc stsdas} package in IRAF. In principle, $W_{\lambda,\mathrm{obs}}$ can be obtained directly by integration, but as can be seen in Figure~\ref{fig:epoch1}, many lines are blended. In some cases one line needs to be fixed in order to measure the potential variation of the other. {\sc ngaussfit} can de-blend up to three lines, with the possibility to keep part of the parameters fixed. This method allows that in spectra with low S/N, an absent or very weak line is sometimes formally best fitted with a line with negative equivalent width (i.e., an emission line, see also Figure~\ref{fig:mosaicNi3F92} and \ref{fig:mosaicFe6D92}); for these lines we obtain upper limits for $W_{\lambda,\mathrm{obs}}$. 
	\par The formal error on the equivalent width is obtained from the error spectrum:
	\begin{equation}
	\sigma W_{\lambda,\mathrm{obs}} = \delta \lambda \sqrt{ \sum_{i}^\mathrm{line}(\sigma_i/F_c)^2},
	\end{equation}
	with $\delta \lambda$ the spectral bin width, $\sigma_i$ the value of the error spectrum in bin $i$ and $F_c$ the continuum flux. The summation runs over two times the full-width-at-half-maximum (FWHM). Furthermore, we add 3\% of the equivalent width to the formal error to account for the uncertainty in placing the continuum. In the low-resolution GMOS spectra, all absorption lines are unresolved and can be fit with a gaussian with a fixed FWHM of  $3.6\,\AA$ (the resolution element), which is determined from the width of the lines in the arc-lamp frames. In xsh we fix $\textrm{FWHM}= 1.0\,\AA$ and $0.7\,\AA$ in the UVB and VIS-arms, respectively, based on the widths of arc lines and telluric emission lines. Some strongly saturated lines (for example the Mg{\,\sc ii} doublet) in these spectra cannot be fit with gaussians and are therefore integrated numerically. The resulting rest frame equivalent widths $W_\lambda=W_{\lambda,\mathrm{obs}} / (1+z_h) $ of the resonance lines in the first two epochs are listed in Table~\ref{tab:COGhost}.\\

	\par To infer column densities of the different ions from the absorption line equivalent widths, we make use of a multi-ion single component curve-of-growth (MISC-COG) analysis \citep{spitzer1978}. 
	The Doppler line width is in principle related to the temperature and the level of turbulence in the absorbing gas. Limited by spectral resolution, it is difficult to distinguish between this and broadening due to the complex physical structure of absorbing clouds in the sight line. Therefore we speak of an 'effective' Doppler parameter, which does not carry information about the temperature of the gas, but is merely a simplification in the model \citep{jenkins1986}. The assumption we need to make in this analysis is that the effective Doppler parameter $b$ is the same for all ion species, and that it is single (one velocity component). In other words, we assume that the velocity structure is the same in all lines, and that it is dominated by small-scale effect such as thermal velocities and turbulence and not by differences in bulk velocities of different gas clouds.

\begin{table*}
\caption{The measured rest-frame equivalent width $W_{\lambda}$ for the resonance lines at the host redshift, and the corresponding ion column densities, obtained with curve-of-growth (COG) analysis. The first column lists the ion and the vacuum rest wavelength ($\AA$); the second column gives the oscillator strength $f_\lambda$. Columns (3) and (4) give the rest-frame $W_\lambda$ in \AA\, measured in epoch1 and epoch2 (GMOS) respectively. For lines for which we have a good measurement both in epoch1 and epoch2, the weighted average is used; otherwise we use the  value measured in epoch1. The values that are finally used for the COG analysis are listed in Column (5). Column (6) gives the best fitting column density for the ion, fitted to a COG with effective Doppler parameter $b=22.1^{+1.7}_{-1.6} \textrm{ km s}^{-1}$. Columns (7) and (8) give the relative abundances with respect to iron and zinc using solar values from \citet{asplund2009} (for notation see Equation~\ref{eq:abunrat}). The references for the wavelength and oscillator strength are in column (9). 
\label{tab:COGhost}}
\begin{tabular}{l@{  }ll r@{ }c@{ }l r@{ }c@{ }l r@{ }c@{ }l cccc} 
\hline
\multicolumn{2}{c}{Line}	&\multicolumn{1}{c}{$f_\lambda$}& \multicolumn{3}{c}{$W_{\lambda}$/\AA}& \multicolumn{3}{c}{$W_{\lambda}/\AA$}& \multicolumn{3}{c}{$<W_{\lambda}/\AA>$}& $\log(N_{\textrm{ion}}/\textrm{cm}^{-2})$ &$\left[ \textrm{X}/\textrm{Fe} \right]$&$\left[ \textrm{X}/\textrm{Zn} \right]$&ref\\
\multicolumn{2}{c}{}	&\multicolumn{1}{c}{}& \multicolumn{3}{c}{epoch1}& \multicolumn{3}{c}{epoch2}& \multicolumn{3}{c}{}&  &&&\\

\hline
Al\,{\sc ii}	&	1670.7874	&	1.740	&	0.418	& $\pm$&	0.093	&	0.338	& $\pm$&	0.098	&	0.380	&$\pm$&		0.067	&	$13.57^{+0.41}_{-0.32}$&$-\,0.61\pm0.42$	&$-\,1.84\pm0.42$&1\\	
&&&&&&&&&&&&&&&\\
Al\,{\sc iii} &        1854.7164 &       0.539	&       0.113 	& $\pm$&    0.054 	&       0.162 	& $\pm$& 0.060 	&   0.135 		& $\pm$&		0.040  	&     $13.02^{+0.13}_{-0.16} $ &$^a$&$^a$&1\\ 
Al\,{\sc iii} &        1862.7895 &       0.268 	&      0.044 	& $\pm$&    0.051  	&       0.113  	& $\pm$& 0.056 	&   0.075 		& $\pm$&   	0.038 	&      ...&& &1\\ 
&&&&&&&&&&&&&&&\\

Si\,{\sc ii}	&	1808.013	&	$2.080 \times 10^{-3}$&	0.292	& $\pm$&	0.062	&	0.267	& $\pm$&	0.068	&	0.281	&$\pm$&		0.046	&	$15.96^{+0.18}_{-0.16} $	&$+\,0.72\pm0.19$&$-\,0.51\pm0.19$&2\\	
&&&&&&&&&&&&&&&\\

Zn\,{\sc ii}	&	2026.136$^b$	&	$5.010 \times 10^{-1}$	&	0.362	& $\pm$&	0.046	&	0.280	& $\pm$&	0.053	&	0.327	&$\pm$&		0.035	&	$13.52^{+0.07}_{-0.07} $	&$+\,1.23\pm0.10$&&3\\							
Zn\,{\sc ii}	&	2062.664	&	$2.460 \times 10^{-1}$	&	0.191	& $\pm$&	0.041	&	0.206	& $\pm$&	0.048	&	0.197	&$\pm$&		0.031	&	...	&&&3\\
&&&&&&&&&&&&&&&\\

Cr\,{\sc ii}	&	2056.2539	&	$1.030 \times 10^{-1}$	&	0.150	& $\pm$&	0.040	&	0.151	& $\pm$&	0.046	&	0.151	&$\pm$&		0.030	&	$13.73^{+0.09}_{-0.10} $ &$+\,0.36\pm0.12$& $-\,0.87\pm0.12$&3	\\							
Cr\,{\sc ii}	&	2066.161	&	$5.120 \times 10^{-2}$	&	0.096	& $\pm$&	0.038	&	0.111	& $\pm$&	0.045	&	0.102	&$\pm$&		0.029	&	...	&&&3\\											
&&&&&&&&&&&&&&&\\
Fe\,{\sc ii}	&	2249.8754$^c$	&	$2.190 \times 10^{-3}$	&	0.129	& $\pm$&	0.034	&	0.151	& $\pm$&	0.040	&	0.139	&$\pm$&		0.026	&	$15.23^{+0.07}_{-0.08}$	&& $-\,1.23\pm0.10$&4\\ 		

Fe\,{\sc ii}	&	2260.7793	&	$2.620 \times 10^{-3}$	&	0.152	& $\pm$&	0.034	&	0.168	& $\pm$&	0.040	&	0.159	&$\pm$&		0.026	&	...	&&&4\\	
Fe\,{\sc ii}	&	2586.6495	&	$7.094 \times 10^{-2}$	&	0.705 	&$\pm$&   0.044	&	0.727 	& $\pm$	&0.048	&	0.715	&$\pm$&	         0.032	&	...	&&&4\\		
&&&&&&&&&&&&&&&\\

Mn\,{\sc ii}	&	2576.877	&	$3.610 \times 10^{-1}$	&	0.265	& $\pm$&	0.032	&	0.257	& $\pm$&	0.037	&	0.262	&$\pm$&		0.024	&	$13.23^{+0.04}_{-0.05}$&$+\,0.07\pm0.09$&$-\,1.16\pm0.04$&	5\\							
Mn\,{\sc ii}	&	2594.499	&	$2.800 \times 10^{-1}$	&	0.206	& $\pm$&	0.030	&	0.207	& $\pm$&	0.034	&	0.207	&$\pm$&		0.023	&	...	&&&5\\							
Mn\,{\sc ii}	&	2606.462	&	$1.980 \times 10^{-1}$	&	0.164	& $\pm$&	0.064	&	$^d$		& 	&			&	0.164	& $\pm$&		0.064	&	...	&&&5\\							
&&&&&&&&&&&&&&&\\
						
Mg\,{\sc ii}	&	2796.352	&	$6.123 \times 10^{-1}$	&	&&	&		&      &			&	0.920	&$\pm$	&	0.147$^e$		&	$15.06^{+0.83}_{-0.60}$	&$-\,0.27\pm0.83$&$-\,1.50\pm0.83$&6\\							
Mg\,{\sc ii}	&	2803.531	&	$3.054 \times 10^{-1}$	&	&&	&		& 	&			&	0.918	&$\pm$	&	0.126$^e$		&	...	&&&6\\	
&&&&&&&&&&&&&&&\\
				
Ca\,{\sc ii}	&	3934.777	&	$6.500 \times 10^{-1}$	&	&&	&		&      &			&	0.907	&$\pm$	&	0.220$^e$		&	$13.74^{+0.47}_{-0.50}$	&$-\,0.33\pm0.50$&$-\,1.56\pm0.50$&7\\							
Ca\,{\sc ii}	&	3969.591	&	$3.220 \times 10^{-1}$	&	&&	&		& 	&			&	0.836	&$\pm$	&	0.185$^e$		&	...	&&&7\\	
\hline
\multicolumn{16}{l}{$^a$The quantities $\left[\textrm{Al \sc iii}/\textrm{Fe} \right]$ and $\left[\textrm{Al \sc iii}/\textrm{Zn} \right]$ do not have any physical meaning, since the Al\,{\sc iii} is probably mainly from a different region than Al\,{\sc ii}, Fe\,{\sc ii} and} \\
\multicolumn{16}{l}{Zn\,{\sc ii} \citep[see also][]{savaglio2004}.}\\
\multicolumn{16}{l}{$^b$The contamination by Cr\,{\sc ii}\,$\lambda2026$ is determined from the other Cr\,{\sc ii} lines and is negligible. The contribution of Mg\,{\sc i}\,$\lambda2026$ cannot be determined, but we}\\
\multicolumn{16}{l}{assume it is small based on the absence of Mg\,{\sc i}\,$\lambda1827$.} \\
\multicolumn{16}{l}{$^c$Fe\,{\sc ii} 2344, 2374, 3282 and 2600 are also clearly detected, but those are blended with fine-structure lines and therefore not used in the abundance study.}\\
\multicolumn{16}{l}{Estimates of their equivalent widths can be found in the appendix.} \\
\multicolumn{16}{l}{$^d$The equivalent width of the line is fixed in the second epoch in order to measure the varied strength of the blended fine-structure line(s).}\\
\multicolumn{16}{l}{$^e$Absorption lines are outside the GMOS range, measurement is from xsh spectrum.}\\
\multicolumn{16}{l}{References for atomic data: (1) NIST Atomic Spectra Database, (2) \citet{bergeson1993b}, (3) \citet{bergeson1993a} (4) \citet{verner1999}, }\\
\multicolumn{16}{l}{(5) \citet{morton2003}, (6) \citet{verner1996}, (7) \citet{morton1991}.}

\end{tabular}
\end{table*}

	\subsubsection{Host galaxy column densities}
	\label{sec:hostgalaxy}
	We use three unblended Fe{\,\sc ii} resonance lines ($\lambda2249$, $\lambda2260$ and $\lambda2586$) to determine $b$ for the line-of-sight gas in the host galaxy (see Table~\ref{tab:COGhost}). Because these lines cover both the linear and the flat part of the COG, both $b$ and the column density for Fe{\,\sc ii}, $N_\mathrm{Fe{\,\sc II}}$, can be constrained. They are dependent variables, therefore fit them simultaneously with a grid method, yielding $b=22.1^{+1.7}_{-1.6}$\,km\,s$^{-1}$ and $\log N_\mathrm{Fe{\,\sc II}}/\mathrm{cm}^{-2}=15.23^{+0.07}_{-0.08}$.	
		\par
\begin{figure}   
	\includegraphics[width=8.5cm]{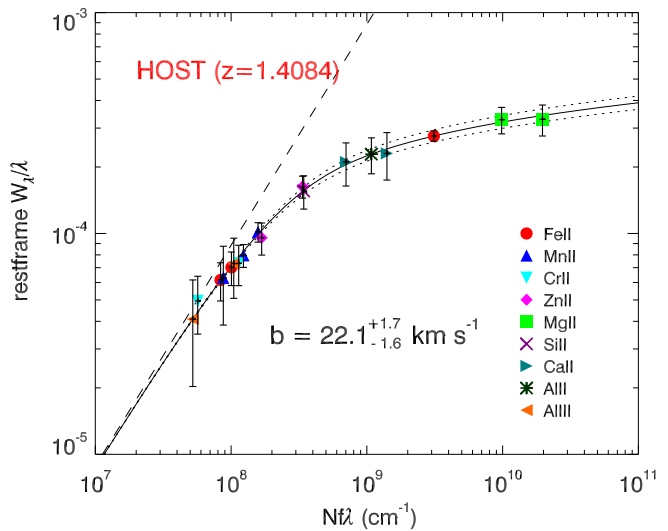}
	\caption{Curve-of-growth (COG) for the resonance lines in the host galaxy of \thisgrb. The solid line gives the best fit COG with $b=22.1\,\textrm{km s}^{-1}$,  the dotted lines show the effect of the error on $b$. The dashed line, which coincides with the linear part of the COG, is a COG for $b=\infty$, that can be used as an approximation for weak lines. }
	\label{fig:COG_host}
	\end{figure}

 The COG with the best fit value for $b$ is shown in Figure~\ref{fig:COG_host}. The ion species other than iron have line transitions either on the linear or on the flat part of the COG; therefore, $b$ has to be assumed to be the same for all ion species in order to determine the column densities. The resulting column densities of the ions are listed in Table~\ref{tab:COGhost}. For most lines we have taken an average of the first two epochs, except Mg{\,\sc ii} and Ca{\,\sc ii} which can only be measured in xsh. The equivalent widths of the Al{\,\sc ii} lines are slightly larger in epoch1 while those of Al{\,\sc iii} are larger in epoch2. While not very significant, it may be a sign of ionisation due to the GRB afterglow \citep[see e.g.,][]{vreeswijk2013}. Due to the low significance we decided not to take this into account in the excitation modelling in Section~\ref{sec:varyfinestruc}.\\
	
	\par Spectral resolution limits the ability to distinguish different velocity components in absorption lines: an instrumental $\textrm{FWHM}=3.6\,\AA\,$(GMOS spectra) corresponds to $\sim200\,\textrm{km s}^{-1}$.
	In xsh-VIS, a velocity component difference of $70\,\textrm{km s}^{-1}$ would be detectable, but we do not see indications for more than one velocity component in any of the lines. This could be due to the low S/N of xsh, because most intermediate to high-resolution spectra of GRBs show absorption lines with at least two components (but see e.g., \citealt{ledoux2009}). 
\par \citet{prochaska2006b} warns that column densities derived with the MISC-COG method are systematically underestimated, especially when the underlying line structure is that of separated clumps of gas at different velocity shifts. In this case one derives a high $b$ with MISC-COG analysis, and therefore a lower column density. Results for a single component COG analysis with  $b\gtrsim 20\,\textrm{km s}^{-1}$ are highly suspect, according to \citet{prochaska2006b}. That COG analysis of a low-resolution spectrum can be accurate is shown by \citet{d'elia2011}, who compare low and high-resolution spectra for the afterglow of GRB\,081008. The COG analysis gives the same result as line fits to the high-resolution spectrum within $3\sigma$. They conclude that this is linked to the low level of saturation of the lines. The equivalent widths of the low-ionisation lines in \thisgrb\,are found to be in the lowest 10\% of the sample of 69 low-resolution spectra conducted by \citet{deugartepostigo2012}, which strengthens the case that a COG analysis on this spectrum can provide accurate results. However, column densities resulting from COG analysis always need to be treated with caution.\\

\begin{figure}   
	\includegraphics[width=8.5cm]{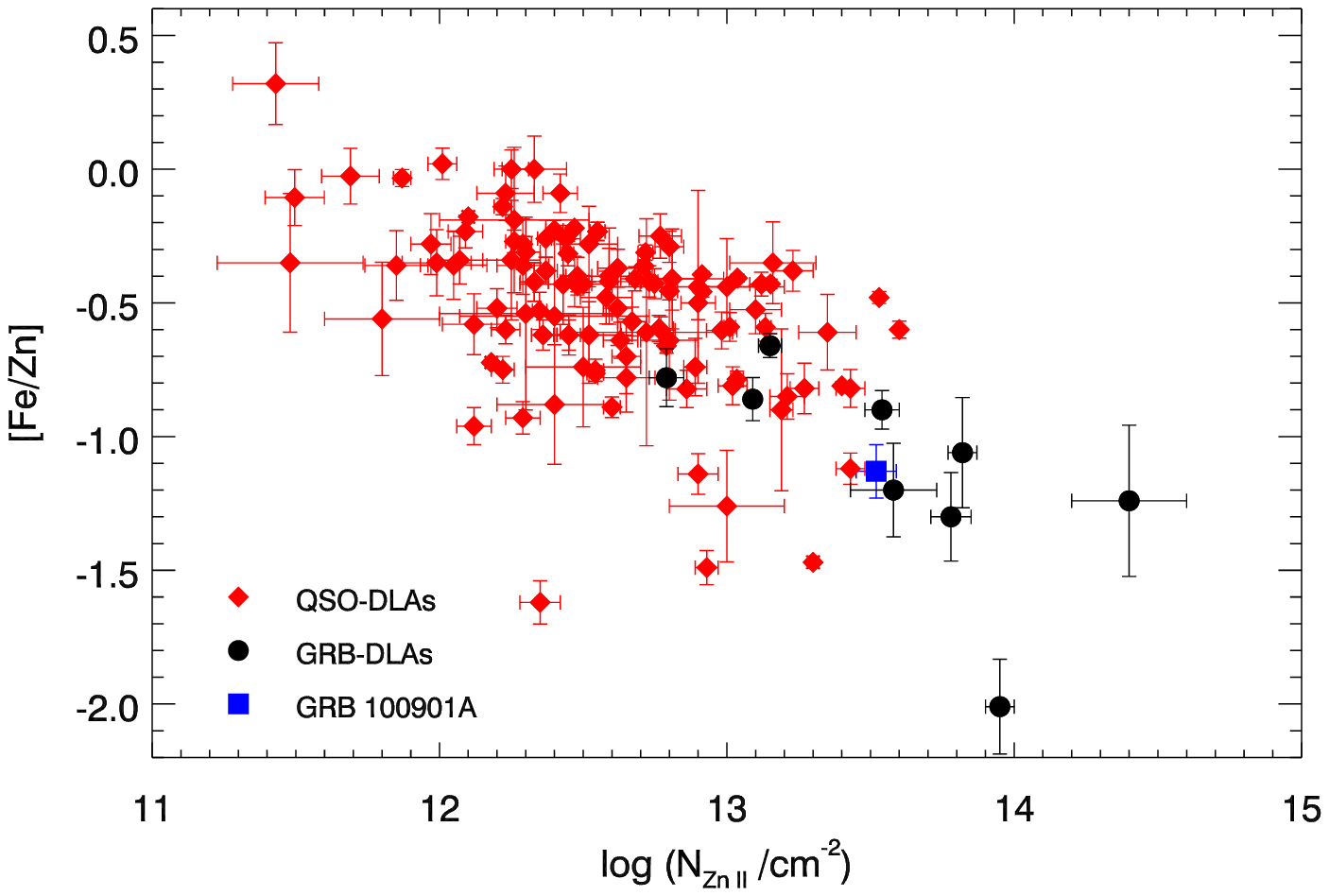}
	\caption{$\left[ \textrm{Fe}/\textrm{Zn} \right]$ as a function of the Zn{\,\sc ii} column density in QSO-DLAs (red diamonds), GRB-DLAs (black circles) and \thisgrb\,(blue square). QSO-DLA values are from \citet{lu1996,prochaska1996,prochaska1997,pettini1999,pettini2000,prochaska2000,prochaska2001}. GRB-DLA values are from \citet{savaglio2003,fynbo2002,fiore2005,watson2006,penprase2006,vreeswijk2007,d'elia2009b,d'elia2011}.}
	\label{fig:depletionsample}
\end{figure}	

\subsubsection{Dust depletion}
	\par The strength of the absorption lines in the spectrum are a measure of the abundance of an element in the gas phase only. Apart from the overall gas-to-dust ratio, the fraction of a specific element that is locked onto dust depends on the element species and on the nature of the dust. Therefore, the pattern of relative abundances, the depletion pattern, might give insight into the amount and nature of the dust. In Table~\ref{tab:COGhost} we list the abundance ratios with respect to the refractory (depleted) element Fe and the non-refractory element Zn, compared to the ratio in a solar abundance environment. We use the following, commonly used notation
	\begin{equation}
	\label{eq:abunrat}
	\left[ \mathrm{X}/\mathrm{Y} \right] \equiv  \log \left(\frac{N\left(\textrm{X}\right)}{N\left(\mathrm{Y}\right)}\right)  - \log \left(\frac{n\left(\mathrm{X}\right)}{n\left(\mathrm{Y}\right)}\right)_{\sun},
	\end{equation}
	with $N\left(\textrm{X}\right)$ the column density of element $X$, and $n\left(\textrm{X}\right)$ the number density of element $X$ in a solar environment \citep{asplund2009}. Differences from zero indicate dust content. We thus assume that relative abundances (both in gas and dust) are solar. Fe, Zn and Cr are iron peak elements, which makes this a reasonable assumption, although the main production pathway of Zn is not fully identified \citep{umeda2002}. Discrepancies in relative abundances might also be due to a different star-formation history than the Milky Way.
	\par In Figure~\ref{fig:depletionsample} we show $\left[ \textrm{Fe}/\textrm{Zn} \right]$ as a function of the Zn column density for the host of \thisgrb, compared with a sample of other GRB-DLAs and QSO-DLAs\footnote{DLA: Damped Lyman-$\alpha$ system: a sight line absorber with $N_\mathrm{H{\,\sc I}}>2 \times 10^{20} $~cm$^{-2}$ \citep{wolfe2005}. A DLA system in a GRB afterglow spectrum is usually due to the host galaxy.}. The relative abundance values indicate a stronger depletion (i.e., more dust) in GRB-DLAs than in QSO-DLAs, but among the GRB-DLAs, \thisgrb\,is not a particularly special case. Both GRB-DLAs and QSO-DLAs appear to follow the overall trend that systems with stronger Zn column densities show stronger depletion. The fact that the two classes populate different regions in the diagram is mainly a sight line effect. QSOs probe galaxies in random orientations with only a small chance that the densest region of a galaxy is intersected, while sight lines to GRB afterglows probe the gas towards the progenitor region, so they will in general show high column densities \citep[see also][]{savaglio2006,prochaska2007,fynbo2009}. Furthermore, for GRB-DLAs with relatively low column densities and weak absorption lines, $\left[ \textrm{Fe}/\textrm{Zn} \right]$ is not well constrained (i.e., not reported in literature) because the fading afterglow prevents building up S/N over several nights of observations. This is not the case for QSO-DLAs, which is why these objects also populate the low $N_{\mathrm{Zn\,{\sc II}}}$ part of the parameter space.	\\
	
\par A more detailed analysis of the dust content can be performed by comparing the heavy-element dust depletion pattern of \thisgrb\,with observations in the interstellar medium (ISM) of the Milky Way \citep{savage1996}. We use the method described in \citet{savaglio2004}. We consider the observed depletion patterns in the MW sight lines trough a warm halo (WH), warm disk and halo (WDH), warm disk (WD) and cool disk (CD) as models. For the fit, we use two free parameters. One is the dust-to-metal ratio relative to the Galactic values $\kappa = \kappa_\mathrm{GRB}/\kappa_J$ ($J$ is one of the 4 depletion patterns). The second is proportional to the metallicity (not known because $N_{\mathrm{H \sc \,I}}$ is not measured in \thisgrb). The two parameters are scaled until they reproduce the observed heavy-element depletion pattern best (minimum $\chi^2$). 
\par
\begin{figure}   
	\includegraphics[width=8.5cm]{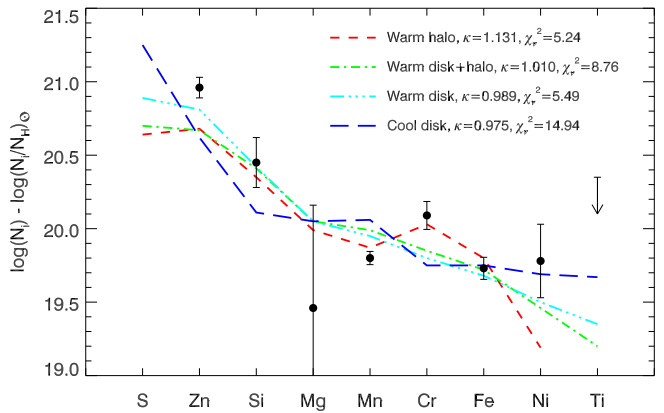}
	\caption{Depletion pattern for the host of \thisgrb, compared with different components in the Milky Way \citep{savage1996}, which are observed patterns that can be adjusted by changing the dust-to-metal ratio $\kappa$. The column density of Ni{\sc\,ii} is not directly measured but follows from the modelling of the fine-structure and metastable levels, see Section~\ref{sec:uvpump}. We include a 3$\sigma$ upper limit for Ti{\sc\,ii} based on the absence of the doublet at a rest wavelength of 1910\,\AA.}
	\label{fig:depletionp}
	\end{figure}

 Results using measurements of 7 elements are shown in Figure~\ref{fig:depletionp}. The WH depletion pattern gives the smallest $\chi^2$ among the 4 patterns, but the WD pattern is also a reasonable fit. The dust-to-metals ratio is between the MW value and 10\% higher. Ni and Mg are the most uncertain measurements. The Mg {\sc ii} column density is likely a lower limit because it suffers from very strong saturation. The Ni {\sc ii} column density is not measured from the resonance lines directly, but follows from the modelling of the time evolution of the Ni {\sc ii}\,$^4F_{9/2}$ metastable level (see Section~\ref{sec:uvpump} and Table~\ref{tab:model}). Excluding these two elements from the fit does not change our results much.
\par For the observed column of metals, and assuming a rate of visual extinction per column density of metals like in the MW \citep{bohlin1978}, the expected optical extinction in GRB100901A would be $A_V  \sim 0.5$. This is higher than the one derived from the spectral energy distribution (SED) of GRB100901A ($A_V =0.21$, Gomboc et al., in prep). This inconsistency is  typically observed in GRB sight lines \citep{schady2011} and indicates that the rate of extinction in the ISM of GRB hosts per column of metals is different from what is observed in the MW.

	\begin{figure}   
	\includegraphics[width=8.5cm]{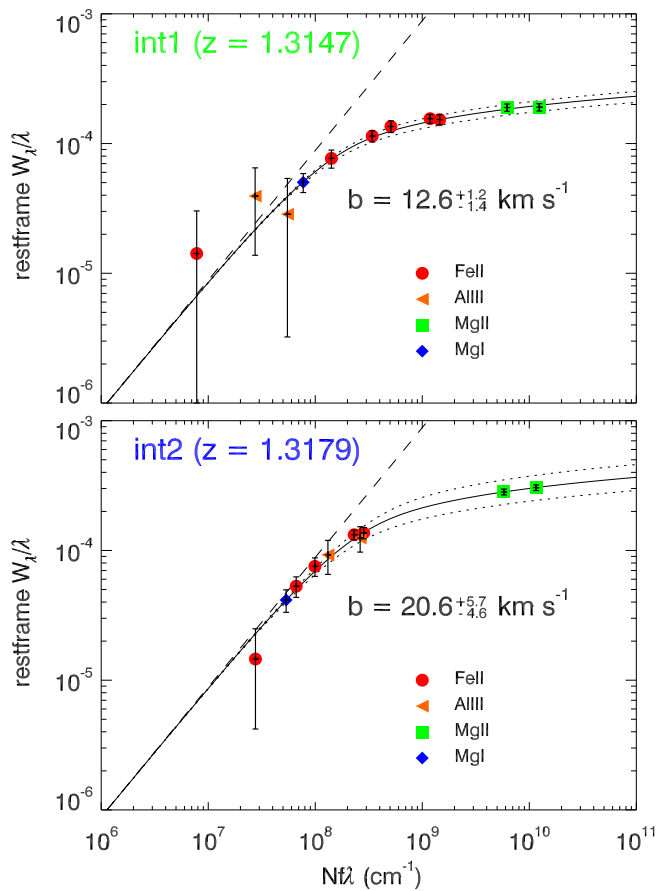}
	\caption{Curve-of-growth for the two intervening systems in the sightline to \thisgrb. The lines used and the best-fit column densities are listed in Table~\ref{tab:COGint}.}
	\label{fig:COG_int}
	\end{figure}

	\subsubsection{Intervening absorbers}
	\label{sec:interv}
	Two intervening absorbers are detected: int1 at $z_1=1.3147$ and int2 at $z_2=1.3179$. If these two systems are not physically associated with each other, and the redshift difference is dominated by cosmological expansion, the velocity difference corresponds to a co-moving separation of 3.9\,Mpc. If the absorbers belong to one system, the velocity difference is 414\,km\,s$^{-1}$, and could be due to motion of galaxies within a cluster. 
	\par Because of the long time span that our observations cover (until long after the jet break, Gomboc et al., in prep), the data set is suited to look for variations in the strength of the lines from the intervening absorbers. This can be interesting in the context of the previously supposed discrepancy between the number of  strong Mg{\,\sc ii} absorbers ($W_\lambda(2796)>1\AA$) in QSO and GRB sight lines; in the latter the redshift number density was found to be about two times as high (\citealt{prochter2006,vergani2009,cucchiara2009}. However, \citet{cucchiara2012} used a larger sample and did not reproduce this discrepancy.
	\par We do not detect variability above the 2$\sigma$ level in any of the two intervening absorbers of \thisgrb, and we see no systematic trend in the ensemble of lines per absorber, nor per ion species. In principle, the non-variation of the equivalent widths of the intervening absorber resonance lines gives a lower limit on the size of the absorbing clouds (i.e., the scale on which the absorbing gas is homogeneous). If the projected apparent size of the afterglow became larger than the projected size of the intervening absorber, the absorption lines from this system would become weaker since part of the light would reach the observer unabsorbed. We use the description of the apparent size of the Blandford-McKee spherical expansion described by \citet{granot1999}, where we use $E_\mathrm{iso} = 6.3 \times 10^{52}$ erg \citep{gorbovskoy2012}. We assume the close circum-burst medium to have a constant density $n=1\mathrm{\,cm}^{-3}$. At the time of the last epoch (7.027 days), we find the source to be $9.6\times10^{16}$\,cm. This gives a lower limit on the size of each of the absorbing clouds of $0.03$\,pc. Although this is not a very meaningful lower size-limit for a gas cloud, we point out that in general it is difficult to put limits on absorber sizes, but see \citet{d'elia2010a} for methods applied to GRB absorbers and \citet{petitjean2000,ellison2004,balashev2011} for examples of size estimates of QSO line-of-sight objects. \\
		
	\par Because the lines of the intervening systems do not vary in strength, we can average the equivalent widths measured in epoch1 and epoch2. The MISC-COG analysis can be applied, because we detect many clear lines from these systems. Figure~\ref{fig:COG_int} shows the best fitting COGs for both intervening absorbers. We carry out the same approach as for the host, by first constraining $b$ with the Fe\,{\sc ii} transitions. We find $b=12.6^{+1.2}_{-1.4}\textrm{ km s}^{-1}$ for int1 and $b=20.6^{+5.7}_{-4.6}\textrm{ km s}^{-1}$ for int2. The equivalent widths and the column densities of each ion species are listed in Table~\ref{tab:COGint}. \\
	
\begin{table*}
\caption{The rest-frame equivalent widths $W_{\lambda}$ for the lines from two intervening systems int1 and int2 at respectively $z_1=1.3147$ and $z_2=1.3179$. $<W_{\lambda}/\AA>$  is the weighted average of the rest frame equivalent width (in \AA ngstrom) measured in GMOS epoch1 and epoch2. The COGs for the two systems are shown in Figure~\ref{fig:COG_int}. The column densities per ion species have been calculated with the COG with the best fitting $b$; the errors on the column densities are the result of taking into account the range in $b$. Especially for lines on the flat part of the COG (e.g., Mg\,{\sc ii}), the effect on $N$ due to $b$ is large.
\label{tab:COGint}}
\begin{tabular}{l @{ }l l r@{ }l@{ }l c r@{ }   l@{ }l c} 
\hline
&&&\multicolumn{4}{c}{int1  $(z_1=1.3147)$}&\multicolumn{4}{c}{int2 $(z_2=1.3179)$}\\
&&&\multicolumn{4}{c}{$b=12.6^{+1.2}_{-1.4}\textrm{ km s}^{-1}$}&\multicolumn{4}{c}{$b=20.6^{+5.7}_{-4.6}\textrm{ km s}^{-1}$}\\
\multicolumn{2}{c}{Line}	&\multicolumn{1}{c}{$f_\lambda$}& \multicolumn{3}{c}{$<W_{\lambda}/\AA>$}& $\log(N_{\textrm{ion}}/\textrm{cm}^{-2})$& \multicolumn{3}{c}{$<W_{\lambda}/\AA>$}& $\log(N_{\textrm{ion}}/\textrm{cm}^{-2})$\\
\hline
Al\,{\sc iii} &        1854.7164 &       0.539 	&   	0.053	&	$\pm$	&	0.046  		&     $12.74^{+0.29}_{-0.43}$	&     	0.229	&	$\pm$	&	0.049	&     $13.42^{+0.27}_{-0.17}$ \\ 
Al\,{\sc iii} &        1862.7895 &       0.268  	&      	0.073	&	$\pm$	&	0.047 		&     ...					&       0.171	&	$\pm$	&	0.049	&      ... \\ 
&&&&&&&&&&\\
Fe\,{\sc ii}	&	2249.8754 &	$2.190 \times 10^{-3}$	&	0.032	&	$\pm$	&	0.036	 	&     $14.28^{+0.15}_{-0.09} $&			&	$^a$		&			&		\\ 
Fe\,{\sc ii}	&	2344.2129	&	$1.252 \times 10^{-1}$	&	0.317	&	$\pm$	&	0.030 		&     	...			&	0.176	&	$\pm$	&	0.027	&	$13.57^{+0.09}_{-0.08}$	\\  
Fe\,{\sc ii}	&	2374.4604	&	$3.297 \times 10^{-2}$	&	0.182	&	$\pm$	&	0.028	 	&     	...			&	0.034	&	$\pm$	&	0.024	&	...	\\    
Fe\,{\sc ii}	&	2382.7641	&	$3.432 \times 10^{-1}$	&	0.365	&	$\pm$	&	0.030	 	&     	...			&	0.325	&	$\pm$	&	0.029	&	...	\\	
Fe\,{\sc ii}	&	2586.6495&	$7.094 \times 10^{-2}$   	&	0.296	&	$\pm$	&	0.027	 	&     	...			&	0.136	&	$\pm$	&	0.023	&	...	\\	
Fe\,{\sc ii}	&	2600.1722&	$2.422 \times 10^{-1}$	&	0.405	&	$\pm$	&	0.028	 	&     	...			&	0.344	&	$\pm$	&	0.027	&	...	\\	
&&&&&&&&&&\\
Mg\,{\sc ii}	&	2796.352	&	$6.123 \times 10^{-1}$	&	0.532	&	$\pm$	&	0.030	 	&     $14.86^{+0.26}_{-0.47}$&	0.845	&	$\pm$	&	0.035	&	$14.83^{+1.39}_{-0.68}$	\\
Mg\,{\sc ii}	&	2803.531	&	$3.054 \times 10^{-1}$	&	0.531	&	$\pm$	&	0.030		&     ...			&	0.792	&	$\pm$	&	0.034	&	...	\\
&&&&&&&&&&\\
Mg\,{\sc i}	&	2852.9642&	1.83   	&	0.143	&	$\pm$	&	0.023		&     $12.17^{+0.09}_{-0.11}$&	0.118	&	$\pm$	&	0.022	&	$12.01^{+0.09}_{-0.11}	$	\\	\hline
\multicolumn{11}{l}{$^a$this line is blended with Ni\,{\sc ii} $^4F_{9/2}\,\lambda2166.23$ at the host redshift.}\\

\end{tabular}
\end{table*}
	
	\begin{figure*}
	\begin{center}
	\resizebox{163.0mm}{!}{\includegraphics{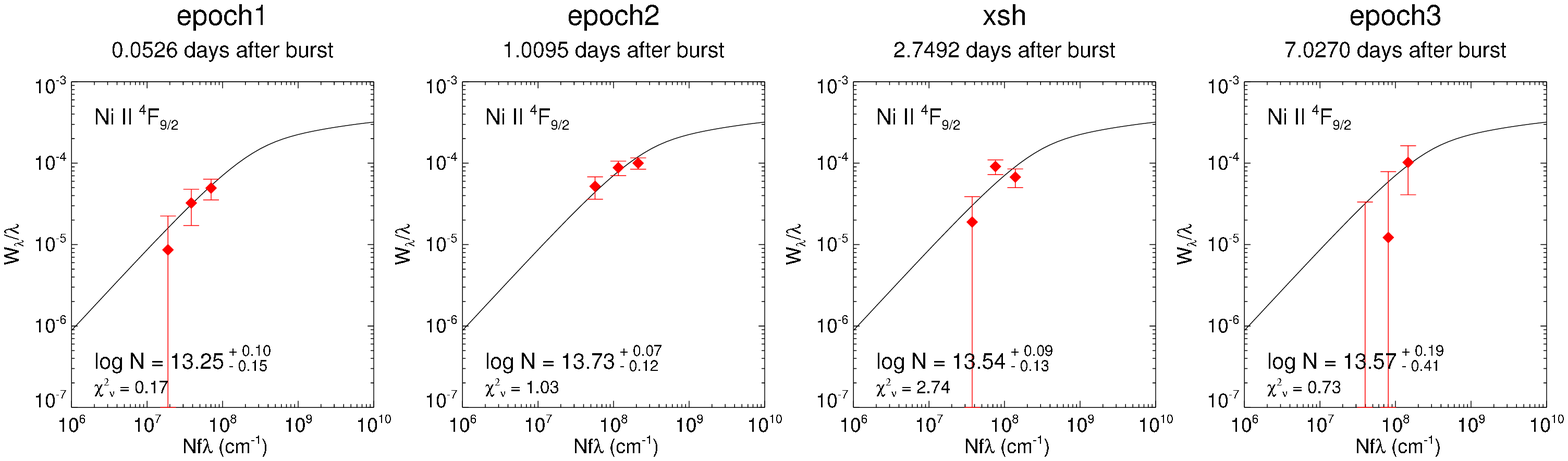}}
\resizebox{40mm}{!}{\includegraphics{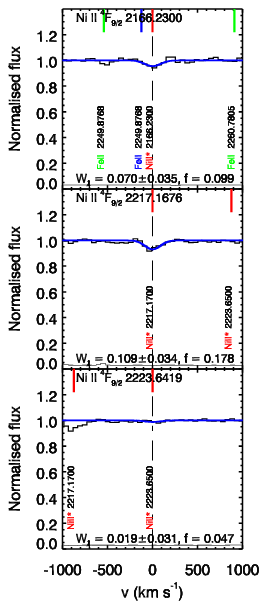}}
\resizebox{40mm}{!}{\includegraphics{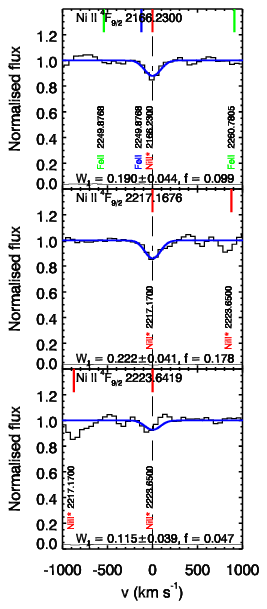}}
\resizebox{40mm}{!}{\includegraphics{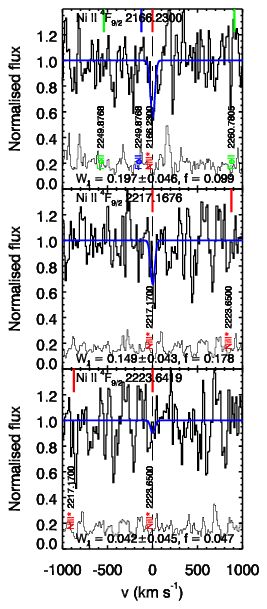}}
\resizebox{40mm}{!}{\includegraphics{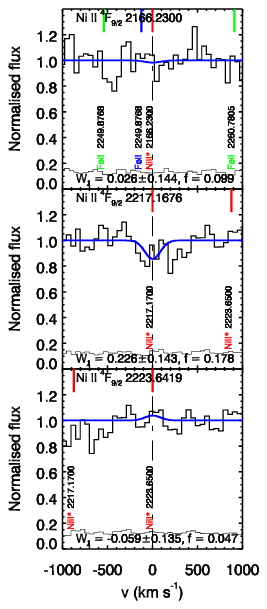}}
\end{center}
\caption{Curves-of-growth (COG) and spectrum excerpts for the three detected lines from the Ni\,{\sc ii} $^4F_{9/2}$ metastable level. Every column represents an epoch, with the observed time since the burst indicated on top. The upper plot shows the COG for $b=22.1\textrm{ km s}^{-1}$ to which we fit the equivalent widths for the three transitions (red diamonds). The best-fit column density $N_{\mathrm{level}}$ is indicated, together with the $\chi^2_{\nu}$ of the COG-fit. Below the COG, the three individual lines are displayed on a velocity scale, with the corresponding gaussian line profile fit. The rest frame equivalent width and oscillator strength are indicated. We use the same line label colouring as in Figure~\ref{fig:epoch1}.}
\label{fig:mosaicNi3F92}
\end{figure*}

	\subsection{Detection and variability of transitions arising from excited levels of Fe\,{\sc ii} and Ni\,{\sc ii}.}
	\label{sec:varyfinestruc}
	At the redshift of the host galaxy we detect lines from fine-structure levels of Fe\,{\sc ii} ($^6D_{7/2}$, $^6D_{5/2}$, $^6D_{3/2}$ and $^6D_{1/2}$) and metastable levels of Fe\,{\sc ii} ($^4F_{9/2}$ and $^4D_{7/2}$) and Ni\,{\sc ii} ($^4F_{9/2}$). The equivalent width of many of these lines varies with time. This provides the opportunity to derive the distance between the burst location and the absorbing material, assuming that the population of these excited states is due to the UV-radiation of the afterglow (see Section~\ref{sec:uvpump}). With high-resolution spectra it would be possible to directly measure the column densities of the ions that are in the excited states, via Voigt profile fits to the fine-structure lines \citep[see e.g.,][]{d'elia2007,vreeswijk2007}. In order to obtain column densities from the equivalent widths, we need to take an intermediate step with the COG, where we fix $b$ to the value found from the resonance Fe{\,\sc ii} lines (see Section~\ref{sec:hostgalaxy}); i.e., we assume that the ions in the excited states are in the same absorbing clouds as the ions in the ground state; for the implications of this assumption see the end of Section~\ref{sec:uvpump}. Note that the value of $b$ will only have a very small influence on the column densities of the excited ions, because the lines are all weak and lie mostly on the linear part of the COG.
	
	Tables~\ref{tab:Fetrans}, \ref{tab:Fetrans2}, \ref{tab:Fetrans_metastab} and \ref{tab:Nitrans_metastab} in the Appendix give an overview of all transitions from the fine-structure and metastable levels of Fe{\,\sc ii} and Ni{\,\sc ii} that would in principle be observable in the GMOS spectral range at this redshift. Due to the low spectral resolution, many of these lines are blended such that the individual variation of the components cannot be measured. The lines that are used in the modelling are shown in boldface in Tables~\ref{tab:Fetrans} to \ref{tab:Nitrans_metastab}.
	
\begin{figure}
	\begin{flushright}
	\resizebox{84.mm}{!}{\includegraphics{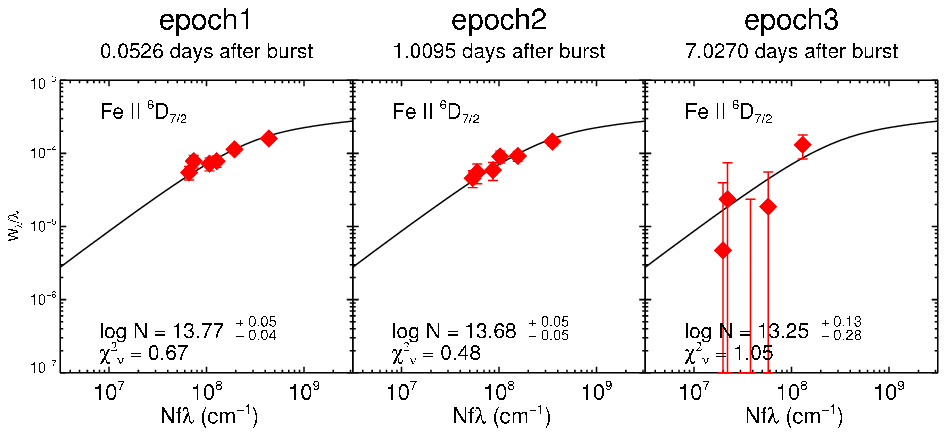}}\\
	\resizebox{84.mm}{!}{\includegraphics{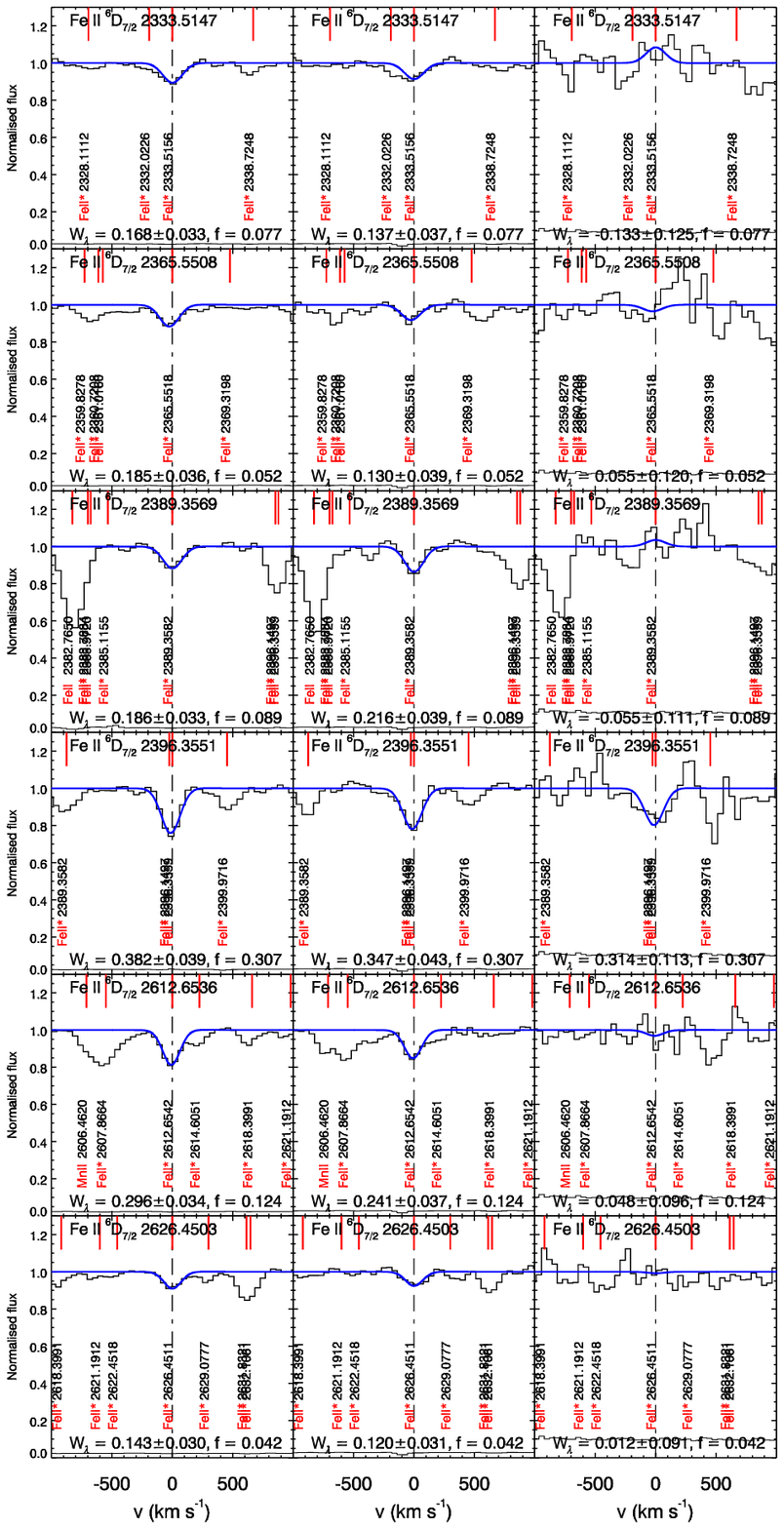}}
	\end{flushright}
\caption{The same as Figure~\ref{fig:mosaicNi3F92}, only for the six detected lines from the Fe\,{\sc ii} $^6D_{7/2}$ fine-structure level. We omit the xsh epoch, because for this level none of the lines was significantly detected due to the low S/N.}
\label{fig:mosaicFe6D92}
\end{figure}

	In order to obtain the column density $N_{\mathrm{level}}$ of ions in a specific excited state in an epoch, the equivalent widths of the transitions arising from this excited level are placed on the COG by assigning an $N_{\mathrm{level}}$ to this level. The best-fit $N_{\mathrm{level}}$ is the value for which the ensemble of transitions fit the COG best (minimisation of the $\chi^2$). Figures~\ref{fig:mosaicNi3F92}  and \ref{fig:mosaicFe6D92} show the COG of the Ni\,{\sc ii}\,$^4F_{9/2}$ and the Fe\,{\sc ii} $^6D_{7/2}$ level per epoch (columns), together with the observed transitions from this level and gaussian fits to the line profiles. The best-fit $N_{\mathrm{level}}$ and its errors are found with a Monte Carlo simulation, which we apply as follows. In every iteration, for each transition an equivalent width is randomly picked from a normal distribution with as mean the measured value and as sigma the measured error. For this simulated ensemble of equivalent widths, the best-fit $N_{\mathrm{level}}$ is determined by minimising the $\chi^2$. After 10\,000 iterations, the distribution of best-fit $N_{\mathrm{level}}$'s is fit with an asymmetric gaussian (i.e., a normal distribution with a different $\sigma$ on both sides), from which we derive the overall best $N_{\mathrm{level}}$ and its lower and upper $1\sigma$ error. Because the fine-structure lines lie mostly on the linear part of the COG, the effect of $b$ is small, and the uncertainty of this value is not taken into account. If two or more transitions of a level are detected in an epoch (i.e., if $|W_\lambda|>\sigma |W_\lambda |$), we obtain a value for $N_{\mathrm{level}}$, otherwise we obtain an upper limit (see Table~\ref{tab:Ns}).

	\begin{table*}

\caption{Lower level column densities $\log \left( N_{\mathrm{level}}(t) / \mathrm{cm}^{-2} \right) $ for Fe{\,\sc ii} and Ni{\,\sc ii} excited states derived from fine-structure line equivalent widths with COG analysis. All errors and limits are 1$\sigma$.
\label{tab:Ns}}
\centering
\begin{tabular}{ll l@{ }l@{ }l@{ }l@{ }l@{ }l@{ }l}
\hline
	&	& \multicolumn{6}{c}{Fe{\,\sc ii}} & \multicolumn{1}{c}{Ni{\,\sc ii}} \\
  \\
&time	(d)& $^6D_{9/2}$ 			& $^6D_{7/2}$ 			      & $^6D_{5/2}$ 			   & $^6D_{3/2}$ 			& $^6D_{1/2}$ 			     & $^4D_{7/2}$ 			& $^4F_{9/2}$  \\ \\
\hline
\\[0.5pt]
epoch1 	& 0.0526 & $15.20^{+0.08}_{-0.09}$ & $13.77^{+0.05}_{-0.04}$ 	& $13.58^{+0.06}_{-0.10}$ & $13.30^{+0.07}_{-0.09}$ & $12.92^{+0.11}_{-0.14}$ 	& $12.54^{+0.11}_{-0.18}$ & $13.25^{+0.10}_{-0.15}$ \\[3.0pt]
epoch2 	& 1.0095 & $15.29^{+0.07}_{-0.10}$ & $13.68^{+0.05}_{-0.05}$ 	& $13.46^{+0.09}_{-0.13}$ & $13.14^{+0.09}_{-0.15}$ & $<12.83$ 			& $12.54^{+0.11}_{-0.24}$ & $13.73^{+0.07}_{-0.12}$ \\[3.0pt]
xsh 		& 2.7492 & $15.16^{+0.12}_{-0.18}$ & $<13.46$				& $<14.02$ 			& $<13.47$ 			& $<13.40$ 			& $<13.03$ 			& $13.54^{+0.09}_{-0.13}$ \\[3.0pt]
epoch3 	& 7.0270 & $15.34^{+0.21}_{-0.48}$ & $<13.37$				& $<13.80$ 			& $<13.39$	 		& $<13.47$ 			& $<12.97$ 			& $<13.75$			\\[3.0pt]

\hline

\end{tabular}
\end{table*}

		\subsubsection{UV-pumping models}
		\label{sec:uvpump}
		
\begin{figure}   
	\includegraphics[width=8.5cm]{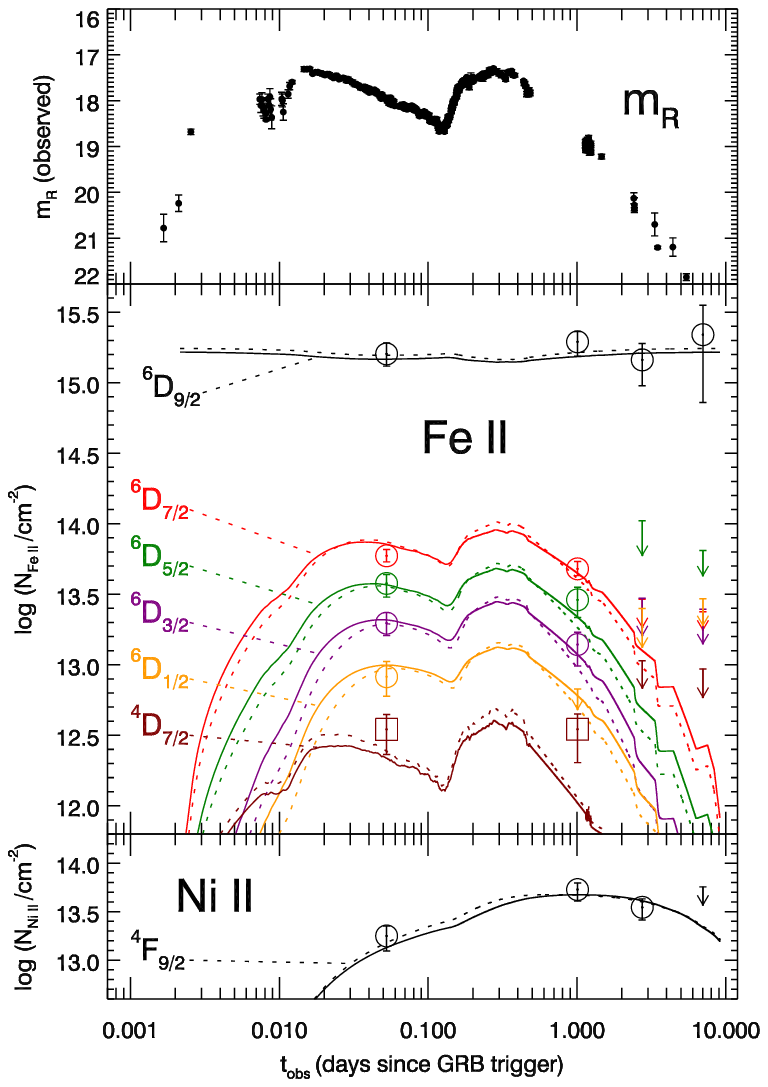}
	\caption{Top panel: observed $R$-band magnitudes from various observatories (see text), which are used as input for the photo-excitation model (see Section~\ref{sec:uvpump}). Middle and bottom panel: column density as a function of observed time for Fe\,{\sc ii} fine-structure and metastable levels, and for the Ni\,{\sc ii}\,$^4F_{9/2}$ metastable level. The open circles and squares (arrows) indicate the column densities (limits) derived for each excited state (see Table~\ref{tab:Ns}). The solid lines are the predicted best fit result of the photo-excitation model with a maximum absorber thickness $l\leq 500$\,pc ($\chi^2_{\nu}=1.15$); the dotted line for a model with fixed absorber thickness $l=1$\,pc ($\chi^2_{\nu}=1.35$), both with $A_V=0.21$.}
	\label{fig:finestruc_variability}
	\end{figure}

		The column densities of the fine-structure and metastable states are found to vary in time (see Table~\ref{tab:Ns} and Figure~\ref{fig:finestruc_variability}). The presence, and variation, of fine-structure lines in GRB afterglow spectra can generally well be explained by excitation due to the UV flux of the GRB afterglow. \citet{prochaska2006a} suggest that this is the dominant excitation mechanism, especially when variability is measured. The early high-resolution spectroscopic data obtained for GRB\,060418 \citep{vreeswijk2007} allowed systematic tests of other excitation mechanisms such as a background IR field and collisional excitation, but also strongly favours the UV-pumping scenario. For \thisgrb\,we assume that UV-pumping is the only relevant photo-excitation mechanism.
		\par We apply the photo-excitation model introduced by \citet{vreeswijk2007,vreeswijk2011,vreeswijk2013}. For technical details we refer to the 2013 paper. The general idea is that the UV-flux of the afterglow temporarily excites ions in a cloud with thickness $l$ at a distance $d$ from the location of the burst. As the afterglow brightness fades in time, the excited levels depopulate after a time which is determined by the Einstein coefficient $A_{ul}$ or, equivalently, the oscillator strength of the transitions considered. For a given set of input parameters, the column density of each excited level as a function of time, $N_{\mathrm{level}}(t)$, is predicted by the model. By comparing this temporal behaviour to the measured $N_{\mathrm{level}}(t)$ for all levels simultaneously, we can optimise the parameters.
		
		 \par The afterglow flux is included as an interpolated series of $R$-band magnitudes\footnote{The observed $R$-band magnitudes come from GCNs \citet{andreev2010a,andreev2010b,andreev2010c,andreev2010e,andreev2010d,kuroda2010b,kuroda2010a,volnova2010}; assembled by Gomboc et al. (in prep.).} $m_R(t)$ (see upper panel Figure~\ref{fig:finestruc_variability}), which corresponds to the rest-frame UV-flux (at $\sim2700\,\AA$) responsible for populating the levels. The monochromatic flux in the host-galaxy rest frame at the GRB-facing side of the absorbing cloud is computed as follows:
\begin{equation}
F_{\nu}^{\mathrm{rest}}(t)=\frac{F_0 \cdot 10^{[{m_R(t)-A_{R,\mathrm{gal}}}]/-2.5}}{1+z_h} \left[ \frac{\lambda_\mathrm{rest}(1+z_h)}{6415\,\AA} \right]^{\beta_\nu} \left[ \frac{D_L}{d}\right]^2
\label{eq:flux}
\end{equation}		 
		in which $F_0=3.02 \times 10^{-20}\,\mathrm{erg\,s}^{-1}\,\mathrm{cm}^{-2}\,\mathrm{Hz}^{-1}$ is the flux of Vega at the effective wavelength ($6410\,\AA$) of the $R$-band \citep{fukugita1995}, $A_{R,\mathrm{gal}}=0.264$ is the Galactic extinction \citep{schlegel1998}, $\lambda_\mathrm{rest}$ is the wavelength array of the relevant transitions at which the flux is required to calculate the amount of excitation, $\beta_\nu$ is the spectral slope and $D_L=9.968\times10^9$\,pc is the luminosity distance. The effective Doppler parameter $b$ is fixed to 22.1 km s$^{-1}$, consistent with how we converted equivalent widths to column densities (see Section~\ref{sec:varyfinestruc}). Parameters that can be constrained from the light curve and/or SED fitting (Gomboc et al., in prep) are kept fixed: optical spectral slope $\beta_\nu=0.82$, and optical extinction in the host galaxy $A_V=0.21$ assuming a Small Magellanic Cloud extinction profile (Gomboc et al., in prep). We performed models with $A_V=0$ as well to study the effect. $A_V$ is not included in Equation~\ref{eq:flux} because the model allows to specify where the optical extinction takes place. We place the extinction in the absorbing cloud, and apply the necessary corrections to the flux. 
		
\begin{table*}

\caption{Photo-excitation modelling results for the environment of \thisgrb~with different settings. Column (1) extinction $A_V$ in the host galaxy, (2) requirement for the distance $d$ from the burst to the absorber, (3) requirement for the thickness $l$ of the absorbing cloud, (4) fit result $d$, (5) fit result $l$, (6) pre-burst column densities of Fe{\,\sc ii} and (7) Ni{\,\sc ii} (ground state) and (8) the reduced chi-square of the fit. See also Figure~\ref{fig:finestruc_variability}.
\label{tab:model}}

\begin{tabular}{lll@{\hspace{1cm}}lllll}
\multicolumn{3}{l}{input}	&							\multicolumn{5}{l}{output}\\
$A_V$	&	$d$	&	$l$ 			&	$d/$pc 		& $l/$pc			 & $\log \left( N_\mathrm{Fe{\,\sc II}}/\mathrm{cm}^{-2}\right)$ 	& $\log \left(N_\mathrm{Ni{\,\sc II}}/\mathrm{cm}^{-2}\right)	$& $\chi^2_{\nu}$\\
\hline
$0.0$	&	free				&	free			&	$250\pm75$ 	& $0^{+240}_{-}$		& $15.25^{+0.05 }_{-0.05}$ 	& $13.91^{+0.07 }_{-0.08}$		& $1.54$ 		\\[3.0pt]	
$0.2$	&	minimal 100 pc	&	free			&	$100^{+61}_{-}$ 	& $1127\pm569 $	& $15.26^{+0.05 }_{-0.06}$	& $14.14^{+0.10 }_{-0.13}$		& $0.73$ 		\\[3.0pt]		
$0.2$	&	free				&	fixed to 1 pc	&	$275\pm12$ 	& 1				& $15.24^{+0.05 }_{-0.06}$ 	& $13.94^{+0.06 }_{-0.07}$		& $1.35$ 		\\[3.0pt]		
$0.2$	&	free				&	fixed to 100 pc	&	$249\pm13$ 	& 100				& $15.24^{+0.05 }_{-0.06}$ 	& $13.95^{+0.06 }_{-0.07}$		& $1.33$ 		\\[3.0pt]	
$0.2$	&	free				&	maximal 500 pc	&$165\pm73$ 	& $500^{+}_{-469}$	& $15.22^{+0.05 }_{-0.06}$ 	& $14.02^{+0.10 }_{-0.13}$		& $1.15$ 		\\[3.0pt]	
\hline
\end{tabular}
\end{table*}

		\par Table~\ref{tab:model} shows the fit results of the model to the measured $N_{\mathrm{level}}(t)$ as listed in Table~\ref{tab:Ns}, with different requirements for the parameters. The model without extinction fits the data best with a small cloud at $\sim250$\,pc, but when we include optical extinction $A_V=0.2$, the data favours a configuration with a very large ($>1$\,kpc) absorbing cloud at a smaller distance. Fixing the cloud size $l$ leads to absorber distances of $d\sim165-275$\,pc. Figure~\ref{fig:finestruc_variability} shows the fit to the data for the model with $l\leq500$\,pc (solid lines) and the model with $l=1$\,pc (dotted lines). Both models are decent fits: our data are not constraining enough to discriminate between the close-large and the distant-small cloud case. But in all cases, the influence of the GRB reaches distances of a few hundreds of parsecs. In Section~\ref{sec:absorberdistance} we will discuss the implications of this estimated absorber distance and compare it to values found in other GRBs.
		\par Though not detectable with the spectral resolution used, it could be possible that ground-state Fe\,{\sc ii} is present in more velocity components than excited Fe\,{\sc ii}, i.e., that there is Fe\,{\sc ii} that is not associated with the excited gas. If there exists more Fe\,{\sc ii} further away, then this would decrease the amount of Fe\,{\sc ii} that is associated with the excited Fe\,{\sc ii}, and hence this would move the absorber closer to the burst, i.e., we would infer a smaller GRB-absorber distance. Very roughly, if this fraction of Fe\,{\sc ii} associated to excited Fe\,{\sc ii} is only 50\% (rather than the now assumed 100\%), then it would decrease the distance by a factor of $\sqrt 2$. So even when assuming such a large fraction (50\%) of the ground-state Fe\,{\sc ii} is further away does not change the inferred distance much.
		
\subsubsection{Atomic data}
\label{sec:discussf}
In the full analysis from the COG to the photo-excitation model fitting, we need to adopt a set of values for vacuum wavelengths $\lambda$ and oscillator strengths $f_{\lambda}$. The results of the analysis described in this paper are obtained with the atomic data given in Tables~\ref{tab:Fetrans}, \ref{tab:Fetrans2}, \ref{tab:Fetrans_metastab} and \ref{tab:Nitrans_metastab}. This atomic parameter set is the same as the one used in {\sc cloudy} \citep{ferland1998}, a widely-used photo-ionisation code, and comes from \citet{verner1999} and references therein \citep[see also ][]{ledoux2009}. The choice for this set is in principle arbitrary, but chosen such that our results can easily be compared with other works on GRB afterglow spectra. We tested the dependence of our results to this input by carrying out the full analysis with a different set of values for $\lambda$ and $f_{\lambda}$. This alternative set is the same as used in \citet{vreeswijk2007} and uses atomic data from \citet{quinet1996}, \citet{morton2003} and \citet{kurucz2003}. This leads to a 2\% lower $b$, and differences up to 0.06 dex in $\log N(t)$. The final resulting $l$ and $d$ of the cloud from the modelling in Section~\ref{sec:uvpump} differ by  $<10$\%, sometimes with lower values for $\chi^2_{\nu}$. This latter fact is, however, not a reason to adopt the alternative set but rather to interpret the scatter in the results as an additional error.

	\subsection{Emission lines}
	\label{sec:emission}
	
	\begin{figure}
\includegraphics[width=8.5cm]{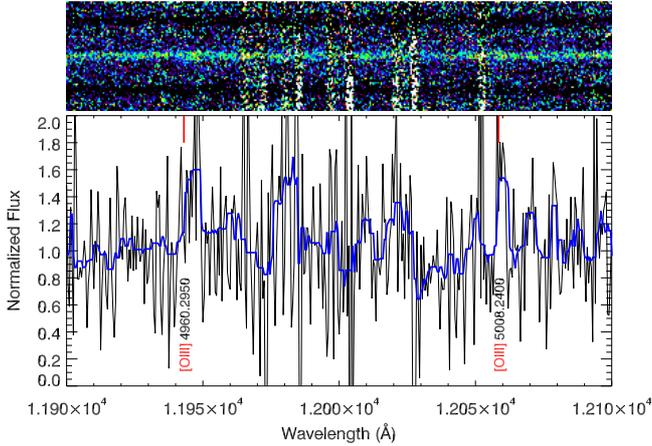}
\caption{\zd\,host emission lines, redshifted to $\sim1.2\mu$ in the xsh-NIR spectrum. The top panel shows the 2D spectrum (0.5\AA/px). The bottom panel shows in black the normalised 1D spectrum (0.5\AA/px); in blue the same spectrum, but with a 10 pixel median filter to make the lines easier to see. The non-identified peaks in the 1D spectrum are due to residuals of skyline subtraction which are visible in the 2D spectrum as vertical white strips. 
\label{fig:xshemission}}
\end{figure}

	Because GRBs originate in star forming galaxies, we have searched for emission lines at the host-galaxy redshift. There is tentative evidence for \zd\,$\lambda5007$ (see Figure~\ref{fig:xshemission}), of which we estimate the flux to be $4.2 \pm 1.3\times10^{-17}\,\textrm{erg s}^{-1}\textrm{ cm}^{-2}$, error including systematics due to flux calibration. \zd\,$\lambda4959$ appears to be present, but is blended with a nearby skyline and cannot be measured. We do not detect H$\alpha$, H$\beta$ or \zt\,due to low S/N and/or skyline blends.

\section{Discussion}
\label{sec:discus}
We have shown that the \thisgrb\, afterglow UV radiation excited Fe{\,\sc ii} and Ni{\,\sc ii} ions in an absorbing cloud located a few hundred parsecs away in the host-galaxy ISM (Section \ref{sec:varyfinestruc}). The ionic column densities we derive for the part of this host galaxy that is along the line of sight (Section \ref{sec:resonancelines}) are very comparable to other GRB hosts where fine-structure lines are detected in the afterglow spectra \citep[see e.g.,][]{penprase2006,prochaska2006a,vreeswijk2007,d'elia2009b}. Furthermore, the part of the galaxy we probe does not appear to be a special case in the context of dust depletion (Section \ref{sec:hostgalaxy}). Modelling of the variation of the lines from the excited states allows us to estimate the distance to, and the size of, the absorbing cloud. This has been possible for about a dozen of bursts; in the following section we will discuss what these absorber distances teach us about the use of GRB afterglows as probes of galaxies. 

\subsection{GRBs as probes of galaxies}
\label{sec:absorberdistance}
GRB afterglows are used as probes of their host galaxies, making use of the fact that the afterglow can be extremely bright for a short time. An advantage (for example over galaxies in quasar sight lines) is that afterglows select star-forming galaxies, under the generally accepted assumption that GRBs are linked to (massive) star formation \citep{woosley1993,paczynski1998}. In order to judge how useful they are as galaxy probes, it is important to understand what the influence of the GRB and its afterglow is on the environment, and what part of the host galaxy we are probing. For about a dozen GRB afterglows (see Table~\ref{tab:otherburst_finestruc_var}), the distance between the region where the absorption lines originate and the site where the burst has gone off has been derived with spectroscopy. The typical absorber distance is of the order of a few hundred pc, so the values we find for \thisgrb\,appear to be normal (Table~\ref{tab:otherburst_finestruc_var}). These values are also in agreement with lower limits on the absorber distance ($>50-100$\,pc) revealed by the presence of Mg{\,\sc i} absorption lines \citep{prochaska2006a}. 
\par Evidently, a GRB is able to affect the ISM out to distances of 100 to 1000\,pc, which can be much further out than only the star-forming region of the GRB progenitor. For example, the largest star-forming region in the Local Group, the Tarantula nebula, has a diameter of $\sim200$\,pc \citep{lebouteiller2008}. Firstly, this means that the GRB and its afterglow could have considerably influenced the environment it probes. Secondly, the absorption lines that we observe are probably not from the immediate star-forming region of the progenitor of the burst. This is important, as many stellar evolution models that are proposed as long GRB progenitors critically rely on a low metallicity \citep[in particular iron group element abundances, see e.g.,][]{woosley2006}. Knowing the distance to the abundance-providing cloud gives more quantitative input into the reliability of the found abundances as tracers of the progenitor abundances.

\subsection{Can low-$R$ spectra be of use in the analysis of fine-structure line variation?}
\label{sec:modellinglowres}
Our analysis of the fine-structure line variability in \thisgrb\, differs from other studies because it is predominantly based on low-resolution spectra, while most other absorber distances are determined from spectra with a $25-40$ times higher spectral resolution (see Table~\ref{tab:otherburst_finestruc_var}). Low resolution allowed us to monitor the spectrum over a much longer time-span. We note that \thisgrb\,remained particularly bright even after a week. The combination of our observations and the excitation modelling shows that we are detecting the peak of the Ni{\,\sc ii} $^4F_{9/2}$ level population at around 1 day after the burst in the observer's frame  (see Figure~\ref{fig:finestruc_variability}), which to date has not been observed. The downside is that low-resolution spectroscopy yields less accurate results, because additional steps and assumptions are required to translate equivalent widths to column densities. Furthermore, it is difficult to disentangle different velocity components, limiting the degree to which we can map the internal dynamical structure of the host galaxy. 
\par When a (multi-epoch) low-resolution data set is available, we have shown that it is worthwhile to measure the variation and estimate the absorber distance by fitting a photo-excitation model to it. Especially the lines from the Ni{\,\sc ii} $^4F_{9/2}$ metastable level have proven to be useful for these kind of long time-span spectroscopic data-sets. These lines are strong, relatively isolated from other common lines in the spectrum, but still close to each other (with GMOS, one would be able to catch all four of them if the burst is at $0.8 <z <1.9$), and it takes longer for the levels to depopulate compared to the Fe{\,\sc ii} fine-structure levels. In terms of recommendations for observing strategies, it would still be most useful to obtain a high-resolution spectrum as soon as possible after the burst, but late-time (after a few days depending on the brightness) spectroscopic follow-up at low resolution can provide additional constraints on the line variation. The information about the different velocity components and their Doppler parameters obtained from the earliest spectrum can be used to interpret the late time low-resolution data more accurately. 

\begin{table*}
\caption{Bursts for which an absorber distance has been derived from (the variability of) lines from excited and/or ionised states in the afterglow spectrum.
Column (1) Burst ID, column (2) number of epochs, column (3) approximate time (observer's frame) between the different epochs, column (4) telescope and spectrograph with which the data have been taken, columns (5) and (6) the ions and levels/lines for which variability is measured and/or modelled, column (7) redshift of the burst, column (8) neutral hydrogen column density columns (9) and (10) distance between the burst and the region where the lines originate, and size of this region following from modelling of the variability, column (11) reference. 
\label{tab:otherburst_finestruc_var}}
\begin{tabular}{l@{ }ll  ll@{ }l l l l@{ }l l} 
Burst 		& \# 		& $\Delta t_{\mathrm{obs}}$ 	& instrument 	& ion &levels/lines							& $z_h$ 	& $\log N_{\mathrm{H{\sc I}}} $ & $d/$pc 		& $l/$pc & ref. \\
\hline
020813 	& 2  		& $\sim16$h 	& KECK/LRIS 			&Fe{\,\sc ii}	& $^6D_{7/2}$ $\lambda$2396  						& 1.25 	& 		-			& $50-100$ 	& - & 1\\ 
 		&   		&  	 		&  VLT/UVES			&			& 	& &&&& \\
050730 	& 2  		& $\sim1$h  	&  VLT/UVES			&Fe{\,\sc ii} 	&$^6D_{7/2}$, $^6D_{5/2}$,$^6D_{3/2}$,$^6D_{1/2}$, 	& 3.97 	& 	22.10	& $124 \pm 20$ 	& $147^{+68}_{-54}$ & 2, 3, 4\\ 
 		&   		&  	 		&   		&			&$^4F_{9/2}$, $^4F_{7/2}$,$^4F_{5/2}$,$^4F_{3/2}$, 		&	 	&&&& \\
 		&   		&  	 		&  					&			&$^4D_{7/2}$, $^4D_{5/2}$ 	& &&&& \\
051111	&1		&			& KECK/HIRES			&Fe{\,\sc ii}	&$^6D_{7/2}$, $^6D_{5/2}$,$^6D_{3/2}$,$^6D_{1/2}$,		&1.55	&- & a few times $10^2$&-& 5, 6\\
		&		&			& 					&Si{\,\sc ii}	&$^2P^{\circ}_{3/2}$		&	& 	 &&&\\	
060206	&$2^a$	&			& WHT/ISIS			&Si{\,\sc ii}	&$^2P^{\circ}_{3/2}$,   							&4.05	& 	$20.85\pm0.1$ 			&$\sim10^3$ &-&7, 8\\	
		&		&			& 				&O{\,\sc i}		&$^3P^{\circ}_{0}$, $^3P^{\circ}_{1}$,&	& 	 &&&\\	
		&		&			& 					&C{\,\sc ii}		&$^2P^{\circ}_{3/2}$&	& 	 &&&\\	
060418	& 6  		& $5-30$ min  	&  VLT/UVES			&Fe{\,\sc ii}	&$^6D_{7/2}$, $^6D_{5/2}$,$^6D_{3/2}$,$^6D_{1/2}$, 	&1.49 	& -		& $480\pm 56$ 	& - &9, 3 \\
 		&   		&  	 		&  	 				&			& $^4F_{9/2}$, $^4D_{7/2}$, 	& &&&& \\
 		&   		&  	 		& 	 				&Ni{\,\sc ii} 	& $^4F_{9/2}$ 				& &&&& \\
080310	&$2-4$	&$10-20$ min	&VLT/UVES			& Fe{\,\sc ii}	&$^6D_{7/2}$, $^6D_{5/2}$,$^6D_{3/2}$,$^6D_{1/2}$,		&2.43 	&$18.70\pm0.1$ 	& $200-400$  & $0-100$& 10, 11\\	
 		&   		&  	 		&  				&			& $^4F_{9/2}$, $^4D_{7/2}$, 	& &&&& \\	
		&		&			&					& Fe{\,\sc iii}	&$^5D$, $^7S_{3}$,&&&&&\\		
		&		&			&					& Si{\,\sc ii}	&$^2P^{\circ}_{3/2}$,&&&&&\\		
		&		&			&					& C{\,\sc ii}	&$^2P^{\circ}_{3/2}$&&&&&\\		
080319B	& 3 		& $40-60$ min  	&  VLT/UVES			&Fe{\,\sc ii} 	&$^6D_{7/2}$, $^6D_{5/2}$,$^6D_{3/2}$,$^6D_{1/2}$, 	& 0.94 	& -				& $560-1700$ 	& - & 12, 3 \\ 
 		&   		&  	 		&  				&			&$^4F_{9/2}$, $^4F_{7/2}$, 	& &&&& \\
 		&   		&  	 		&  					&			&$^4D_{7/2}$, $^4D_{5/2}$, 	& &&&& \\
080330	&1		& 			& VLT/UVES			&Fe{\,\sc ii} 	&$^6D_{7/2}$, $^6D_{5/2}$,$^6D_{3/2}$,$^6D_{1/2}$, 	&1.51	&-		&$79^{+11}_{-14}$&-& 13, 3 \\		
		&		&			&				& 			&$^4F_{9/2},$&&&&\\		
		&		&			&					& Si{\,\sc ii}	&$^2P^{\circ}_{3/2}$,&&&&&\\		
 		&   		&  	 		&  &Ni{\,\sc ii} 	& $^4F_{9/2}$ 				& &&&& \\
081008	&$4^{a}$	&$7-25$ min	&VLT/UVES			& Fe{\,\sc ii}	&$^6D_{7/2}$, $^6D_{5/2}$,$^6D_{3/2}$,				&1.97	&$21.11 \pm 0.10$	&$52 \pm 6$  and $ 200^{+60}_{-80}$&-&14\\		
		&		& 			&		& 			&$^4F_{9/2}$, $^4D_{7/2}$, &&&&\\		
		&		&			&	VLT/FORS2				& Si{\,\sc ii}	&$^2P^{\circ}_{3/2}$,&&&&&\\		
		&   		&  	 		&  				&Ni{\,\sc ii} 	& $^4F_{9/2}$ 				& &&&& \\
090926	&$4^{a}$	&			&VLT/X-shooter			&Fe{\,\sc ii}	&$^6D_{7/2}$, $^4F_{9/2}$,						&2.11	& $21.60 \pm 0.07$ 	& $677\pm42$ and $5\times10^3$ &-&15, 3\\ 
		&		&			&					&Si{\,\sc ii}	&$^2P^{\circ}_{3/2}$,&&&&&\\	
		&		&			& 					&O{\,\sc i}		&$^3P^{\circ}_{0}$, $^3P^{\circ}_{1}$,&	& 	 &&&\\	
		&		&			& 					&C{\,\sc ii}		&$^2P^{\circ}_{3/2}$,&	& 	 &&&\\	
 		&   		&  	 		&  					&Ni{\,\sc ii} 	& $^4F_{9/2}$ 				& &&&& \\
100219A	&1		&			& VLT/X-shooter		&Si{\,\sc ii}	&$^2P^{\circ}_{3/2}$	&4.67&$21.14 \pm 0.15$&$300$ and $10^{3}$&& 16\\ 
100901A	& 4  		&  1 h - 1 wk 	&  Gemini-N/GMOS		&Fe{\,\sc ii} 	&$^6D_{7/2}$, $^6D_{5/2}$,$^6D_{3/2}$,$^6D_{1/2}$, 	& 1.41 & 	-		& a few times $10^2$ 	& $1 - 1000 $	&17 \\ 
 		&   		&  	 		&   	&			& $^4D_{7/2}$, 	& &&&& \\
 		&   		&  	 		&  	VLT/X-shooter				&Ni{\,\sc ii} 	& $^4F_{9/2}$ 				& &&&& \\
		\hline
\multicolumn{11}{l}{$^{a}$More than one epoch is obtained, but the distance measurement did not follow from modelling the variation.}\\
\multicolumn{11}{l}{References: (1) \citet{dessauges2006}; (2) \citet{ledoux2009}; (3) \citet{vreeswijk2011}; (4) \citet{d'elia2007}; (5) \citet{penprase2006};  }\\
\multicolumn{11}{l}{(6) \citet{prochaska2006a}; (7) \citet{fynbo2006}; (8) \citet{thone2008}; (9) \citet{vreeswijk2007}; (10) \citet{decia2012}; }\\
\multicolumn{11}{l}{(11) \citet{vreeswijk2013}; (12) \citet{d'elia2009a}; (13) \citet{d'elia2009b}; (14) \citet{d'elia2011}; (15) \citet{d'elia2010b}; }\\
\multicolumn{11}{l}{(16) \citet{thone2012}; (17) this work.}
\end{tabular}
\end{table*}

\section{Summary and conclusions}
\label{sec:conclusions}
We have analysed the optical to near-IR spectra of the afterglow of \thisgrb. Our data set consists of three low-resolution spectra obtained with Gemini-N/GMOS at 1 hour, 1 day and at seven days after the burst, and one medium-resolution VLT/\mbox{X-shooter} spectrum taken nearly 3 days post burst. At a redshift of 1.4084 a wealth of metal resonance and fine-structure lines is detected, allowing us to estimate column densities of Fe, Si, Cr, Mn, Zn, Mg, Ca and Al with a multi-ion single component curve-of-growth (MISC-COG) analysis. The obtained metal column densities are similar to what has been found in many GRB sightlines. Ly-$\alpha$ is not detected, so no metallicity can be determined. 
\par $\left[\textrm{Fe}/\textrm{Zn}\right] =-1.13 \pm 0.10$ suggests the presence of dust in the host galaxy. A comparison of the observed dust depletion pattern with sight lines in the Milky Way \citep{savage1996} reveals a slight preference for a warm-halo-like depletion pattern. 
\par The spectrum shows two intervening absorbers at redshifts 1.3147 and 1.3179. Lines of Fe{\,\sc ii}, Al{\,\sc ii}, Al{\,\sc iii}, Mg{\,\sc i} and Mg{\,\sc ii} have been detected and the corresponding column densities are estimated with the MISC-COG technique. We do not see significant variation in the strength of any of these lines from the intervening absorbers. Under simple assumptions for the projected source size as a function of time, this lack of variability sets a lower limit of 0.03\,pc on the size (or homogeneity scale) of the absorbing clouds in these foreground galaxies.   
\par At the host-galaxy redshift we detect lines arising from fine-structure levels of the ground state of Fe{\,\sc ii} and metastable levels of Fe{\,\sc ii} and Ni{\,\sc ii}. The strengths of these lines are found to vary significantly in time. By grouping transitions that arise from the same lower level, we have used the COG from the Fe{\,\sc ii} resonance lines to retrieve the temporal behaviour of the column density of each of these excited ion species. This information is fed into a photo-excitation model, that uses the rest-frame UV-flux of the burst as input. We assume that the excited levels are populated due to the UV-radiation produced by the afterglow. The model computes the distance between the burst location and the cloud of gas in which the lines originate, and the size of this cloud. For \thisgrb, we estimate an absorber distance of a few hundred pc, which appears to be typical compared with similar studies. 
\par This is the first time that a fine-structure line variability analysis is predominantly based on low-resolution spectra, and as a consequence of that, also the longest time span over which such variation is detected and modelled. The decrease of the Ni\,{\sc ii}\,$^4F_{9/2}$ metastable level population has not been detected before; our data, in combination with our modelling results, show that we cover the peak of the excitation of this level. We argue that applying the fine-structure variability model to low-resolution data can yield a sensible estimate for the absorber distance, but additional intermediate steps (such as the COG) are necessary, which may cause additional (systematic) errors. 

\section*{Acknowledgments}
This research is partly based on observations obtained at the Gemini Observatory, which is operated by the Association Research in Astronomy, Inc., under a cooperative agreement with the NSF on behalf of the Gemini partnership.\\
The anonymous referee is thanked for useful suggestions that helped to significantly improve the paper. We thank H.\,J.~van Eerten for useful discussion on the evolving source size of GRBs, and A.\,J.\,J.~Raassen for a clear introduction to electron configurations, their notations and nomenclature. \\
O.E.H. is funded by NOVA. J.P.U.F. acknowledges support from the ERC-StG grant EGGS-278202. The Dark Cosmology Centre is funded by the DNRF. A.G. acknowledges funding from the Slovenian Research Agency and from the Centre of Excellence for Space Sciences and Technologies SPACE-SI, an operation partly financed by the European Union, European Regional Development Fund and Republic of Slovenia, Ministry of Higher Education, Science and Technology. A.P. is grateful for support of the RFBR grants 12-02-01336-a and 11-01-92202-Mong-a. R.A.M.J. acknowledges support from the ERC via Advanced Investigator Grant no. 247295.

\bibliographystyle{mn2e_fix}
\bibliography{biblio_GRB_abb}

\begin{thebibliography}{95}
\expandafter\ifx\csname natexlab\endcsname\relax\def\natexlab#1{#1}\fi

\bibitem[{{Andreev}, {Sergeev} \& {Pozanenko}(2010{\natexlab{a}}){Andreev},
  {Sergeev}, \& {Pozanenko}}]{andreev2010a}
{Andreev} M., {Sergeev} A., {Pozanenko} A., 2010{\natexlab{a}}, GCN, 11166, 1

\bibitem[{{Andreev}, {Sergeev} \& {Pozanenko}(2010{\natexlab{b}}){Andreev},
  {Sergeev}, \& {Pozanenko}}]{andreev2010b}
---, 2010{\natexlab{b}}, GCN, 11168, 1

\bibitem[{{Andreev}, {Sergeev} \& {Pozanenko}(2010{\natexlab{c}}){Andreev},
  {Sergeev}, \& {Pozanenko}}]{andreev2010c}
---, 2010{\natexlab{c}}, GCN, 11191, 1

\bibitem[{{Andreev}, {Sergeev} \& {Pozanenko}(2010{\natexlab{d}}){Andreev},
  {Sergeev}, \& {Pozanenko}}]{andreev2010e}
---, 2010{\natexlab{d}}, GCN, 11201, 1

\bibitem[{{Andreev} {et~al}\mbox{.}(2010){Andreev}, {Sergeev}, {Pozanenko},
  {Parakhin}, {Velichko}, {Borachok}, \& {Petkov}}]{andreev2010d}
{Andreev} M., {Sergeev} A., {Pozanenko} A., {Parakhin} N., {Velichko} S.,
  {Borachok} N., {Petkov} V., 2010, GCN, 11200, 1

\bibitem[{{Asplund} {et~al}\mbox{.}(2009){Asplund}, {Grevesse}, {Sauval}, \&
  {Scott}}]{asplund2009}
{Asplund} M., {Grevesse} N., {Sauval} A.~J., {Scott} P., 2009, ARAA, 47, 481

\bibitem[{{Balashev} {et~al}\mbox{.}(2011){Balashev}, {Petitjean}, {Ivanchik},
  {Ledoux}, {Srianand}, {Noterdaeme}, \& {Varshalovich}}]{balashev2011}
{Balashev} S.~A., {Petitjean} P., {Ivanchik} A.~V., {Ledoux} C., {Srianand} R.,
  {Noterdaeme} P., {Varshalovich} D.~A., 2011, MNRAS, 418, 357

\bibitem[{{Berger} {et~al}\mbox{.}(2006){Berger}, {Penprase}, {Cenko},
  {Kulkarni}, {Fox}, {Steidel}, \& {Reddy}}]{berger2006}
{Berger} E., {Penprase} B.~E., {Cenko} S.~B., {Kulkarni} S.~R., {Fox} D.~B.,
  {Steidel} C.~C., {Reddy} N.~A., 2006, APJ, 642, 979

\bibitem[{{Bergeson} \& {Lawler}(1993{\natexlab{a}})}]{bergeson1993a}
{Bergeson} S.~D., {Lawler} J.~E., 1993{\natexlab{a}}, APJL, 414, L137

\bibitem[{{Bergeson} \& {Lawler}(1993{\natexlab{b}})}]{bergeson1993b}
---, 1993{\natexlab{b}}, APJ, 408, 382

\bibitem[{{Bohlin}, {Savage} \& {Drake}(1978){Bohlin}, {Savage}, \&
  {Drake}}]{bohlin1978}
{Bohlin} R.~C., {Savage} B.~D., {Drake} J.~F., 1978, APJ, 224, 132

\bibitem[{{Chornock} {et~al}\mbox{.}(2010){Chornock}, {Berger}, {Fox}, {Levan},
  {Tanvir}, \& {Wiersema}}]{chornock2010}
{Chornock} R., {Berger} E., {Fox} D., {Levan} A.~J., {Tanvir} N.~R., {Wiersema}
  K., 2010, GCN, 11164, 1

\bibitem[{{Cucchiara} {et~al}\mbox{.}(2009){Cucchiara}, {Jones}, {Charlton},
  {Fox}, {Einsig}, \& {Narayanan}}]{cucchiara2009}
{Cucchiara} A., {Jones} T., {Charlton} J.~C., {Fox} D.~B., {Einsig} D.,
  {Narayanan} A., 2009, APJ, 697, 345

\bibitem[{{Cucchiara} {et~al}\mbox{.}(2012){Cucchiara}, {Prochaska}, {Zhu},
  {M{\'e}nard}, {Fynbo}, {Fox}, {Chen}, {Cooksey}, {Cenko}, {Perley}, {Bloom},
  {Berger}, {Tanvir}, {D'Elia}, {Vergani}, {Lopez}, {Chornock}, \&
  {deJaeger}}]{cucchiara2012}
{Cucchiara} A. {et~al.}, 2012, ArXiv e-prints

\bibitem[{{De Cia} {et~al}\mbox{.}(2012){De Cia}, {Ledoux}, {Fox}, {Vreeswijk},
  {Smette}, {Petitjean}, {Bj{\"o}rnsson}, {Fynbo}, {Hjorth}, \&
  {Jakobsson}}]{decia2012}
{De Cia} A. {et~al.}, 2012, A\&A, 545, A64

\bibitem[{{de Ugarte Postigo} {et~al}\mbox{.}(2012){de Ugarte Postigo},
  {Fynbo}, {Th{\"o}ne}, {Christensen}, {Gorosabel}, {Milvang-Jensen},
  {Schulze}, {Jakobsson}, {Wiersema}, {S{\'a}nchez-Ram{\'{\i}}rez}, {Leloudas},
  {Zafar}, {Malesani}, \& {Hjorth}}]{deugartepostigo2012}
{de Ugarte Postigo} A. {et~al.}, 2012, A\&A, 548, A11

\bibitem[{{de Ugarte Postigo} {et~al}\mbox{.}(2011){de Ugarte Postigo},
  {Th{\"o}ne}, {Goldoni}, {Fynbo}, \& {the X-shooter GRB
  collaboration}}]{deugartepostigo2011}
{de Ugarte Postigo} A., {Th{\"o}ne} C.~C., {Goldoni} P., {Fynbo} J.~P.~U., {the
  X-shooter GRB collaboration}, 2011, Astronomische Nachrichten, 332, 297

\bibitem[{{D'Elia} {et~al}\mbox{.}(2011){D'Elia}, {Campana}, {Covino},
  {D'Avanzo}, {Piranomonte}, \& {Tagliaferri}}]{d'elia2011}
{D'Elia} V., {Campana} S., {Covino} S., {D'Avanzo} P., {Piranomonte} S.,
  {Tagliaferri} G., 2011, MNRAS, 418, 680

\bibitem[{{D'Elia} {et~al}\mbox{.}(2010{\natexlab{a}}){D'Elia}, {Fiore},
  {Goldoni}, {D'Odorico}, {Campana}, {Covino}, {D'Avanzo}, {Meurs}, {Norci}, \&
  {Tagliaferri}}]{d'elia2010a}
{D'Elia} V. {et~al.}, 2010{\natexlab{a}}, MNRAS, 401, 385

\bibitem[{{D'Elia} {et~al}\mbox{.}(2007){D'Elia}, {Fiore}, {Meurs},
  {Chincarini}, {Melandri}, {Norci}, {Pellizza}, {Perna}, {Piranomonte},
  {Sbordone}, {Stella}, {Tagliaferri}, {Vergani}, {Ward}, {Angelini},
  {Antonelli}, {Burrows}, {Campana}, {Capalbi}, {Cimatti}, {Costa}, {Cusumano},
  {Della Valle}, {Filliatre}, {Fontana}, {Frontera}, {Fugazza}, {Gehrels},
  {Giannini}, {Giommi}, {Goldoni}, {Guetta}, {Israel}, {Lazzati}, {Malesani},
  {Marconi}, {Mason}, {Mereghetti}, {Mirabel}, {Molinari}, {Moretti}, {Nousek},
  {Perri}, {Piro}, {Stratta}, {Testa}, \& {Vietri}}]{d'elia2007}
---, 2007, A\&A, 467, 629

\bibitem[{{D'Elia} {et~al}\mbox{.}(2009{\natexlab{a}}){D'Elia}, {Fiore},
  {Perna}, {Krongold}, {Covino}, {Fugazza}, {Lazzati}, {Nicastro}, {Antonelli},
  {Campana}, {Chincarini}, {D'Avanzo}, {Della Valle}, {Goldoni}, {Guetta},
  {Guidorzi}, {Meurs}, {Mirabel}, {Molinari}, {Norci}, {Piranomonte}, {Stella},
  {Stratta}, {Tagliaferri}, \& {Ward}}]{d'elia2009a}
---, 2009{\natexlab{a}}, APJ, 694, 332

\bibitem[{{D'Elia} {et~al}\mbox{.}(2009{\natexlab{b}}){D'Elia}, {Fiore},
  {Perna}, {Krongold}, {Vergani}, {Campana}, {Covino}, {D'Avanzo}, {Fugazza},
  {Goldoni}, {Guidorzi}, {Meurs}, {Norci}, {Piranomonte}, {Tagliaferri}, \&
  {Ward}}]{d'elia2009b}
---, 2009{\natexlab{b}}, A\&A, 503, 437

\bibitem[{{D'Elia} {et~al}\mbox{.}(2010{\natexlab{b}}){D'Elia}, {Fynbo},
  {Covino}, {Goldoni}, {Jakobsson}, {Matteucci}, {Piranomonte}, {Sollerman},
  {Th{\"o}ne}, {Vergani}, {Vreeswijk}, {Watson}, {Wiersema}, {Zafar}, {de
  Ugarte Postigo}, {Flores}, {Hjorth}, {Kaper}, {Levan}, {Malesani},
  {Milvang-Jensen}, {Pian}, {Tagliaferri}, \& {Tanvir}}]{d'elia2010b}
---, 2010{\natexlab{b}}, A\&A, 523, A36

\bibitem[{{Dessauges-Zavadsky} {et~al}\mbox{.}(2006){Dessauges-Zavadsky},
  {Chen}, {Prochaska}, {Bloom}, \& {Barth}}]{dessauges2006}
{Dessauges-Zavadsky} M., {Chen} H.-W., {Prochaska} J.~X., {Bloom} J.~S.,
  {Barth} A.~J., 2006, APJL, 648, L89

\bibitem[{{D'Odorico} {et~al}\mbox{.}(2006){D'Odorico}, {Dekker}, {Mazzoleni},
  {Vernet}, {Guinouard}, {Groot}, {Hammer}, {Rasmussen}, {Kaper}, {Navarro},
  {Pallavicini}, {Peroux}, \& {Zerbi}}]{d'odorico2006}
{D'Odorico} S. {et~al.}, 2006, in Society of Photo-Optical Instrumentation
  Engineers (SPIE) Conference Series, Vol. 6269, SPIE Conference Series

\bibitem[{{El{\'{\i}}asd{\'o}ttir}
  {et~al}\mbox{.}(2009){El{\'{\i}}asd{\'o}ttir}, {Fynbo}, {Hjorth}, {Ledoux},
  {Watson}, {Andersen}, {Malesani}, {Vreeswijk}, {Prochaska}, {Sollerman}, \&
  {Jaunsen}}]{eliasdottir2009}
{El{\'{\i}}asd{\'o}ttir} {\'A}. {et~al.}, 2009, APJ, 697, 1725

\bibitem[{{Ellison} {et~al}\mbox{.}(2004){Ellison}, {Ibata}, {Pettini},
  {Lewis}, {Aracil}, {Petitjean}, \& {Srianand}}]{ellison2004}
{Ellison} S.~L., {Ibata} R., {Pettini} M., {Lewis} G.~F., {Aracil} B.,
  {Petitjean} P., {Srianand} R., 2004, A\&A, 414, 79

\bibitem[{{Ferland} {et~al}\mbox{.}(1998){Ferland}, {Korista}, {Verner},
  {Ferguson}, {Kingdon}, \& {Verner}}]{ferland1998}
{Ferland} G.~J., {Korista} K.~T., {Verner} D.~A., {Ferguson} J.~W., {Kingdon}
  J.~B., {Verner} E.~M., 1998, PASP, 110, 761

\bibitem[{{Fiore} {et~al}\mbox{.}(2005){Fiore}, {D'Elia}, {Lazzati}, {Perna},
  {Sbordone}, {Stratta}, {Meurs}, {Ward}, {Antonelli}, {Chincarini}, {Covino},
  {Di Paola}, {Fontana}, {Ghisellini}, {Israel}, {Frontera}, {Marconi},
  {Stella}, {Vietri}, \& {Zerbi}}]{fiore2005}
{Fiore} F. {et~al.}, 2005, APJ, 624, 853

\bibitem[{{Fukugita}, {Shimasaku} \& {Ichikawa}(1995){Fukugita}, {Shimasaku},
  \& {Ichikawa}}]{fukugita1995}
{Fukugita} M., {Shimasaku} K., {Ichikawa} T., 1995, PASP, 107, 945

\bibitem[{{Fynbo} {et~al}\mbox{.}(2002){Fynbo}, {Gorosabel}, {M{\o}ller},
  {Hjorth}, {Andersen}, {Egholm}, {Jensen}, {Pedersen}, {Thomsen}, \&
  {Weidinger}}]{fynbo2002}
{Fynbo} J.~P.~U. {et~al.}, 2002, in Lighthouses of the Universe: The Most
  Luminous Celestial Objects and Their Use for Cosmology, {Gilfanov} M.,
  {Sunyeav} R., {Churazov} E., eds., p. 187

\bibitem[{{Fynbo} {et~al}\mbox{.}(2009){Fynbo}, {Jakobsson}, {Prochaska},
  {Malesani}, {Ledoux}, {de Ugarte Postigo}, {Nardini}, {Vreeswijk},
  {Wiersema}, {Hjorth}, {Sollerman}, {Chen}, {Th{\"o}ne}, {Bj{\"o}rnsson},
  {Bloom}, {Castro-Tirado}, {Christensen}, {De Cia}, {Fruchter}, {Gorosabel},
  {Graham}, {Jaunsen}, {Jensen}, {Kann}, {Kouveliotou}, {Levan}, {Maund},
  {Masetti}, {Milvang-Jensen}, {Palazzi}, {Perley}, {Pian}, {Rol}, {Schady},
  {Starling}, {Tanvir}, {Watson}, {Xu}, {Augusteijn}, {Grundahl}, {Telting}, \&
  {Quirion}}]{fynbo2009}
---, 2009, APJS, 185, 526

\bibitem[{{Fynbo} {et~al}\mbox{.}(2006){Fynbo}, {Starling}, {Ledoux},
  {Wiersema}, {Th{\"o}ne}, {Sollerman}, {Jakobsson}, {Hjorth}, {Watson},
  {Vreeswijk}, {M{\o}ller}, {Rol}, {Gorosabel}, {N{\"a}r{\"a}nen}, {Wijers},
  {Bj{\"o}rnsson}, {Castro Cer{\'o}n}, {Curran}, {Hartmann}, {Holland},
  {Jensen}, {Levan}, {Limousin}, {Kouveliotou}, {Nelemans}, {Pedersen},
  {Priddey}, \& {Tanvir}}]{fynbo2006}
---, 2006, A\&A, 451, L47

\bibitem[{{Goldoni} {et~al}\mbox{.}(2006){Goldoni}, {Royer}, {Fran{\c c}ois},
  {Horrobin}, {Blanc}, {Vernet}, {Modigliani}, \& {Larsen}}]{goldoni2006}
{Goldoni} P., {Royer} F., {Fran{\c c}ois} P., {Horrobin} M., {Blanc} G.,
  {Vernet} J., {Modigliani} A., {Larsen} J., 2006, in Presented at the Society
  of Photo-Optical Instrumentation Engineers (SPIE) Conference, Vol. 6269, SPIE
  Conference Series

\bibitem[{{Gorbovskoy} {et~al}\mbox{.}(2012){Gorbovskoy}, {Lipunova},
  {Lipunov}, {Kornilov}, {Belinski}, {Shatskiy}, {Tyurina}, {Kuvshinov},
  {Balanutsa}, {Chazov}, {Kuznetsov}, {Zimnukhov}, {Kornilov}, {Sankovich},
  {Krylov}, {Ivanov}, {Chvalaev}, {Poleschuk}, {Konstantinov}, {Gress},
  {Yazev}, {Budnev}, {Krushinski}, {Zalozhnich}, {Popov}, {Tlatov},
  {Parhomenko}, {Dormidontov}, {Senik}, {Yurkov}, {Sergienko}, {Varda},
  {Kudelina}, {Castro-Tirado}, {Gorosabel}, {S{\'a}nchez-Ram{\'{\i}}rez},
  {Jelinek}, \& {Tello}}]{gorbovskoy2012}
{Gorbovskoy} E.~S. {et~al.}, 2012, MNRAS, 421, 1874

\bibitem[{{Granot}, {Piran} \& {Sari}(1999){Granot}, {Piran}, \&
  {Sari}}]{granot1999}
{Granot} J., {Piran} T., {Sari} R., 1999, APJ, 513, 679

\bibitem[{{Guidorzi} {et~al}\mbox{.}(2010){Guidorzi}, {Cano}, {Melandri}, \&
  {Kopac}}]{guidorzi2010}
{Guidorzi} C., {Cano} Z., {Melandri} A., {Kopac} D., 2010, GCN, 11160, 1

\bibitem[{{Hjorth} {et~al}\mbox{.}(2012){Hjorth}, {Malesani}, {Jakobsson},
  {Jaunsen}, {Fynbo}, {Gorosabel}, {Kr{\"u}hler}, {Levan}, {Micha{\l}owski},
  {Milvang-Jensen}, {M{\o}ller}, {Schulze}, {Tanvir}, \& {Watson}}]{hjorth2012}
{Hjorth} J. {et~al.}, 2012, APJ, 756, 187

\bibitem[{{Immler} {et~al}\mbox{.}(2010){Immler}, {Barthelmy}, {Baumgartner},
  {Beardmore}, {Campana}, {D'Elia}, {Evans}, {Gelbord}, {Godet}, {Gronwall},
  {Guidorzi}, {Holland}, {Hoversten}, {Littlejohns}, {Marshall}, {O'Brien},
  {Osborne}, {Pagani}, {Page}, {Palmer}, {Pritchard}, {Rowlinson},
  {Sbarufatti}, {Siegel}, {Stamatikos}, \& {Starling}}]{immler2010a}
{Immler} S. {et~al.}, 2010, GCN, 11159, 1

\bibitem[{{Jenkins}(1986)}]{jenkins1986}
{Jenkins} E.~B., 1986, Astrophyiscal Journal, 304, 739

\bibitem[{{Kuroda} {et~al}\mbox{.}(2010{\natexlab{a}}){Kuroda}, {Hanayama},
  {Miyaji}, {Watanabe}, {Yanagisawa}, {Yoshida}, {Ohta}, \&
  {Kawai}}]{kuroda2010b}
{Kuroda} D., {Hanayama} H., {Miyaji} T., {Watanabe} J., {Yanagisawa} K.,
  {Yoshida} M., {Ohta} K., {Kawai} N., 2010{\natexlab{a}}, GCN, 11205, 1

\bibitem[{{Kuroda} {et~al}\mbox{.}(2010{\natexlab{b}}){Kuroda}, {Yanagisawa},
  {Shimizu}, {Toda}, {Nagayama}, {Yoshida}, {Ohta}, \& {Kawai}}]{kuroda2010a}
{Kuroda} D., {Yanagisawa} K., {Shimizu} Y., {Toda} H., {Nagayama} S., {Yoshida}
  M., {Ohta} K., {Kawai} N., 2010{\natexlab{b}}, GCN, 11172, 1

\bibitem[{{Kurucz}(2003)}]{kurucz2003}
{Kurucz} R.~L., 2003, in IAU Symposium, Vol. 210, Modelling of Stellar
  Atmospheres, {Piskunov} N., {Weiss} W.~W., {Gray} D.~F., eds., p.~45

\bibitem[{{Lebouteiller} {et~al}\mbox{.}(2008){Lebouteiller}, {Bernard-Salas},
  {Brandl}, {Whelan}, {Wu}, {Charmandaris}, {Devost}, \&
  {Houck}}]{lebouteiller2008}
{Lebouteiller} V., {Bernard-Salas} J., {Brandl} B., {Whelan} D.~G., {Wu} Y.,
  {Charmandaris} V., {Devost} D., {Houck} J.~R., 2008, APJ, 680, 398

\bibitem[{{Ledoux} {et~al}\mbox{.}(2009){Ledoux}, {Vreeswijk}, {Smette}, {Fox},
  {Petitjean}, {Ellison}, {Fynbo}, \& {Savaglio}}]{ledoux2009}
{Ledoux} C., {Vreeswijk} P.~M., {Smette} A., {Fox} A.~J., {Petitjean} P.,
  {Ellison} S.~L., {Fynbo} J.~P.~U., {Savaglio} S., 2009, A\&A, 506, 661

\bibitem[{{Lu} {et~al}\mbox{.}(1996){Lu}, {Sargent}, {Barlow}, {Churchill}, \&
  {Vogt}}]{lu1996}
{Lu} L., {Sargent} W.~L.~W., {Barlow} T.~A., {Churchill} C.~W., {Vogt} S.~S.,
  1996, APJS, 107, 475

\bibitem[{{Modigliani} {et~al}\mbox{.}(2010){Modigliani}, {Goldoni}, {Royer},
  {Haigron}, {Guglielmi}, {Fran{\c c}ois}, {Horrobin}, {Bristow}, {Vernet},
  {Moehler}, {Kerber}, {Ballester}, {Mason}, \& {Christensen}}]{modigliani2010}
{Modigliani} A. {et~al.}, 2010, in Society of Photo-Optical Instrumentation
  Engineers (SPIE) Conference Series, Vol. 7737, SPIE Conference Series

\bibitem[{{Morton}(1991)}]{morton1991}
{Morton} D.~C., 1991, APJS, 77, 119

\bibitem[{{Morton}(2003)}]{morton2003}
---, 2003, APJS, 149, 205

\bibitem[{{Paczy\'nski}(1998)}]{paczynski1998}
{Paczy\'nski} B., 1998, APJL, 494, L45

\bibitem[{{Penprase} {et~al}\mbox{.}(2006){Penprase}, {Berger}, {Fox},
  {Kulkarni}, {Kadish}, {Kerber}, {Ofek}, {Kasliwal}, {Hill}, {Schaefer}, \&
  {Reed}}]{penprase2006}
{Penprase} B.~E. {et~al.}, 2006, APJ, 646, 358

\bibitem[{{Perna}, {Lazzati} \& {Fiore}(2003){Perna}, {Lazzati}, \&
  {Fiore}}]{perna2003}
{Perna} R., {Lazzati} D., {Fiore} F., 2003, APJ, 585, 775

\bibitem[{{Petitjean} {et~al}\mbox{.}(2000){Petitjean}, {Aracil}, {Srianand},
  \& {Ibata}}]{petitjean2000}
{Petitjean} P., {Aracil} B., {Srianand} R., {Ibata} R., 2000, A\&A, 359, 457

\bibitem[{{Pettini} {et~al}\mbox{.}(1999){Pettini}, {Ellison}, {Steidel}, \&
  {Bowen}}]{pettini1999}
{Pettini} M., {Ellison} S.~L., {Steidel} C.~C., {Bowen} D.~V., 1999, APJ, 510,
  576

\bibitem[{{Pettini} {et~al}\mbox{.}(2000){Pettini}, {Ellison}, {Steidel},
  {Shapley}, \& {Bowen}}]{pettini2000}
{Pettini} M., {Ellison} S.~L., {Steidel} C.~C., {Shapley} A.~E., {Bowen} D.~V.,
  2000, APJ, 532, 65

\bibitem[{{Prochaska}(2006)}]{prochaska2006b}
{Prochaska} J.~X., 2006, APJ, 650, 272

\bibitem[{{Prochaska}, {Chen} \& {Bloom}(2006){Prochaska}, {Chen}, \&
  {Bloom}}]{prochaska2006a}
{Prochaska} J.~X., {Chen} H., {Bloom} J.~S., 2006, APJ, 648, 95

\bibitem[{{Prochaska} {et~al}\mbox{.}(2007){Prochaska}, {Chen},
  {Dessauges-Zavadsky}, \& {Bloom}}]{prochaska2007}
{Prochaska} J.~X., {Chen} H., {Dessauges-Zavadsky} M., {Bloom} J.~S., 2007,
  ApJ, 666, 267

\bibitem[{{Prochaska} \& {Wolfe}(1996)}]{prochaska1996}
{Prochaska} J.~X., {Wolfe} A.~M., 1996, APJ, 470, 403

\bibitem[{{Prochaska} \& {Wolfe}(1997)}]{prochaska1997}
---, 1997, APJ, 474, 140

\bibitem[{{Prochaska} \& {Wolfe}(2000)}]{prochaska2000}
---, 2000, APJL, 533, L5

\bibitem[{{Prochaska} {et~al}\mbox{.}(2001){Prochaska}, {Wolfe}, {Tytler},
  {Burles}, {Cooke}, {Gawiser}, {Kirkman}, {O'Meara}, \&
  {Storrie-Lombardi}}]{prochaska2001}
{Prochaska} J.~X. {et~al.}, 2001, APJS, 137, 21

\bibitem[{{Prochter} {et~al}\mbox{.}(2006){Prochter}, {Prochaska}, {Chen},
  {Bloom}, {Dessauges-Zavadsky}, {Foley}, {Lopez}, {Pettini}, {Dupree}, \&
  {Guhathakurta}}]{prochter2006}
{Prochter} G.~E. {et~al.}, 2006, APJL, 648, L93

\bibitem[{{Quinet}, {Le Dourneuf} \& {Zeippen}(1996){Quinet}, {Le Dourneuf}, \&
  {Zeippen}}]{quinet1996}
{Quinet} P., {Le Dourneuf} M., {Zeippen} C.~J., 1996, A\&AS, 120, 361

\bibitem[{{Sakamoto} {et~al}\mbox{.}(2010){Sakamoto}, {Barthelmy},
  {Baumgartner}, {Cummings}, {Gehrels}, {Krimm}, {Immler}, {Markwardt},
  {Palmer}, {Stamatikos}, {Tueller}, \& {Ukwatta}}]{sakamoto2010}
{Sakamoto} T. {et~al.}, 2010, GCN, 11169, 1

\bibitem[{{Savage} \& {Sembach}(1996)}]{savage1996}
{Savage} B.~D., {Sembach} K.~R., 1996, ARAA, 34, 279

\bibitem[{{Savaglio}(2006)}]{savaglio2006}
{Savaglio} S., 2006, New Journal of Physics, 8, 195

\bibitem[{{Savaglio}(2012)}]{savaglio2012}
---, 2012, Astronomische Nachrichten, 333, 480

\bibitem[{{Savaglio} \& {Fall}(2004)}]{savaglio2004}
{Savaglio} S., {Fall} S.~M., 2004, APJ, 614, 293

\bibitem[{{Savaglio}, {Fall} \& {Fiore}(2003){Savaglio}, {Fall}, \&
  {Fiore}}]{savaglio2003}
{Savaglio} S., {Fall} S.~M., {Fiore} F., 2003, APJ, 585, 638

\bibitem[{{Savaglio}, {Glazebrook} \& {Le Borgne}(2009){Savaglio},
  {Glazebrook}, \& {Le Borgne}}]{savaglio2009}
{Savaglio} S., {Glazebrook} K., {Le Borgne} D., 2009, APJ, 691, 182

\bibitem[{{Schady} {et~al}\mbox{.}(2011){Schady}, {Savaglio}, {Kr{\"u}hler},
  {Greiner}, \& {Rau}}]{schady2011}
{Schady} P., {Savaglio} S., {Kr{\"u}hler} T., {Greiner} J., {Rau} A., 2011,
  A\&A, 525, A113

\bibitem[{{Schlegel}, {Finkbeiner} \& {Davis}(1998){Schlegel}, {Finkbeiner}, \&
  {Davis}}]{schlegel1998}
{Schlegel} D.~J., {Finkbeiner} D.~P., {Davis} M., 1998, APJ, 500, 525

\bibitem[{{Spitzer}(1978)}]{spitzer1978}
{Spitzer} L., 1978, {Physical processes in the interstellar medium}. New York
  Wiley-Interscience, 1978.~333 p.

\bibitem[{{Th{\"o}ne} {et~al}\mbox{.}(2012){Th{\"o}ne}, {Fynbo}, {Goldoni}, {de
  Ugarte Postigo}, {Campana}, {Vergani}, {Covino}, {Kruehler}, {Kaper},
  {Tanvir}, {Zafar}, {D'Elia}, {Gorosabel}, {Greiner}, {Groot}, {Hammer},
  {Jakobsson}, {Klose}, {Levan}, {Milvang-Jensen}, {Nicuesa Guelbenzu},
  {Palazzi}, {Piranomonte}, {Tagliaferri}, {Watson}, {Wiersema}, \&
  {Wijers}}]{thone2012}
{Th{\"o}ne} C.~C. {et~al.}, 2012, ArXiv e-prints

\bibitem[{{Th{\"o}ne} {et~al}\mbox{.}(2008){Th{\"o}ne}, {Wiersema}, {Ledoux},
  {Starling}, {de Ugarte Postigo}, {Levan}, {Fynbo}, {Curran}, {Gorosabel},
  {van der Horst}, {Llorente}, {Rol}, {Tanvir}, {Vreeswijk}, {Wijers}, \&
  {Kewley}}]{thone2008}
---, 2008, A\&A, 489, 37

\bibitem[{{Umeda} \& {Nomoto}(2002)}]{umeda2002}
{Umeda} H., {Nomoto} K., 2002, APJ, 565, 385

\bibitem[{{Vacca}, {Cushing} \& {Rayner}(2003){Vacca}, {Cushing}, \&
  {Rayner}}]{vacca2003}
{Vacca} W.~D., {Cushing} M.~C., {Rayner} J.~T., 2003, Publications of the ASP,
  115, 389

\bibitem[{{van Paradijs} {et~al}\mbox{.}(1997){van Paradijs}, {Groot},
  {Galama}, {Kouveliotou}, {Strom}, {Telting}, {Rutten}, {Fishman}, {Meegan},
  {Pettini}, {Tanvir}, {Bloom}, {Pedersen}, {N{\o}rdgaard-Nielsen},
  {Linden-V{\o}rnle}, {Melnick}, {van der Steene}, {Bremer}, {Naber}, {Heise},
  {in't Zand}, {Costa}, {Feroci}, {Piro}, {Frontera}, {Zavattini}, {Nicastro},
  {Palazzi}, {Bennett}, {Hanlon}, \& {Parmar}}]{paradijs1997}
{van Paradijs} J. {et~al.}, 1997, Nature, 386, 686

\bibitem[{{Vergani} {et~al}\mbox{.}(2009){Vergani}, {Petitjean}, {Ledoux},
  {Vreeswijk}, {Smette}, \& {Meurs}}]{vergani2009}
{Vergani} S.~D., {Petitjean} P., {Ledoux} C., {Vreeswijk} P., {Smette} A.,
  {Meurs} E.~J.~A., 2009, A\&A, 503, 771

\bibitem[{{Verner} {et~al}\mbox{.}(1996){Verner}, {Ferland}, {Korista}, \&
  {Yakovlev}}]{verner1996}
{Verner} D.~A., {Ferland} G.~J., {Korista} K.~T., {Yakovlev} D.~G., 1996, APJ,
  465, 487

\bibitem[{{Verner} {et~al}\mbox{.}(1999){Verner}, {Verner}, {Korista},
  {Ferguson}, {Hamann}, \& {Ferland}}]{verner1999}
{Verner} E.~M., {Verner} D.~A., {Korista} K.~T., {Ferguson} J.~W., {Hamann} F.,
  {Ferland} G.~J., 1999, APJS, 120, 101

\bibitem[{{Vernet} {et~al}\mbox{.}(2011){Vernet}, {Dekker}, {D'Odorico},
  {Kaper}, {Kjaergaard}, {Hammer}, {Randich}, {Zerbi}, {Groot}, {Hjorth},
  {Guinouard}, {Navarro}, {Adolfse}, {Albers}, {Amans}, {Andersen}, {Andersen},
  {Binetruy}, {Bristow}, {Castillo}, {Chemla}, {Christensen}, {Conconi},
  {Conzelmann}, {Dam}, {De Caprio}, {De Ugarte Postigo}, {Delabre}, {Di
  Marcantonio}, {Downing}, {Elswijk}, {Finger}, {Fischer}, {Flores},
  {Francois}, {Goldoni}, {Guglielmi}, {Haigron}, {Hanenburg}, {Hendriks},
  {Horrobin}, {Horville}, {Jessen}, {Kerber}, {Kern}, {Kiekebusch}, {Kleszcz},
  {Klougart}, {Kragt}, {Larsen}, {Lizon}, {Lucuix}, {Mainieri}, {Manuputy},
  {Martayan}, {Mason}, {Mazzoleni}, {Michaelsen}, {Modigliani}, {Moehler},
  {M{\o}ller}, {Norup S{\o}rensen}, {N{\o}rregaard}, {Peroux}, {Patat}, {Pena},
  {Pragt}, {Reinero}, {Riga}, {Riva}, {Roelfsema}, {Royer}, {Sacco}, {Santin},
  {Schoenmaker}, {Spano}, {Sweers}, {Ter Horst}, {Tintori}, {Tromp}, {van
  Dael}, {van der Vliet}, {Venema}, {Vidali}, {Vinther}, {Vola}, {Winters},
  {Wistisen}, {Wulterkens}, \& {Zacchei}}]{vernet2011}
{Vernet} J. {et~al.}, 2011, ArXiv e-prints

\bibitem[{{Volnova} {et~al}\mbox{.}(2010){Volnova}, {Pozanenko}, E., \&
  {Korobtsev}}]{volnova2010}
{Volnova} A., {Pozanenko} A., E. K., {Korobtsev} I., 2010, GCN, 11270, 1

\bibitem[{{Vreeswijk} {et~al}\mbox{.}(2004){Vreeswijk}, {Ellison}, {Ledoux},
  {Wijers}, {Fynbo}, {M{\o}ller}, {Henden}, {Hjorth}, {Masi}, {Rol}, {Jensen},
  {Tanvir}, {Levan}, {Castro Cer{\'o}n}, {Gorosabel}, {Castro-Tirado},
  {Fruchter}, {Kouveliotou}, {Burud}, {Rhoads}, {Masetti}, {Palazzi}, {Pian},
  {Pedersen}, {Kaper}, {Gilmore}, {Kilmartin}, {Buckle}, {Seigar}, {Hartmann},
  {Lindsay}, \& {van den Heuvel}}]{vreeswijk2004}
{Vreeswijk} P.~M. {et~al.}, 2004, A\&A, 419, 927

\bibitem[{{Vreeswijk} {et~al}\mbox{.}(2001){Vreeswijk}, {Fruchter}, {Kaper},
  {Rol}, {Galama}, {van Paradijs}, {Kouveliotou}, {Wijers}, {Pian}, {Palazzi},
  {Masetti}, {Frontera}, {Savaglio}, {Reinsch}, {Hessman}, {Beuermann},
  {Nicklas}, \& {van den Heuvel}}]{vreeswijk2001}
---, 2001, APJ, 546, 672

\bibitem[{{Vreeswijk} {et~al}\mbox{.}(2013){Vreeswijk}, {Ledoux}, {Raassen},
  {Smette}, {De Cia}, {Wo{\'z}niak}, {Fox}, {Vestrand}, \&
  {Jakobsson}}]{vreeswijk2013}
---, 2013, A\&A, 549, A22

\bibitem[{{Vreeswijk} {et~al}\mbox{.}(2007){Vreeswijk}, {Ledoux}, {Smette},
  {Ellison}, {Jaunsen}, {Andersen}, {Fruchter}, {Fynbo}, {Hjorth}, {Kaufer},
  {M{\o}ller}, {Petitjean}, {Savaglio}, \& {Wijers}}]{vreeswijk2007}
---, 2007, A\&A, 468, 83

\bibitem[{{Vreeswijk} {et~al}\mbox{.}(2011){Vreeswijk}, {Ledoux}, {Smette},
  {Ellison}, {Jaunsen}, {Andersen}, {Fruchter}, {Fynbo}, {Hjorth}, {Kaufer},
  {M{\o}ller}, {Petitjean}, {Savaglio}, \& {Wijers}}]{vreeswijk2011}
---, 2011, A\&A, 532, C3

\bibitem[{{Watson} {et~al}\mbox{.}(2006){Watson}, {Fynbo}, {Ledoux},
  {Vreeswijk}, {Hjorth}, {Smette}, {Andersen}, {Aoki}, {Augusteijn},
  {Beardmore}, {Bersier}, {Castro Cer{\'o}n}, {D'Avanzo}, {Diaz-Fraile},
  {Gorosabel}, {Hirst}, {Jakobsson}, {Jensen}, {Kawai}, {Kosugi}, {Laursen},
  {Levan}, {Masegosa}, {N{\"a}r{\"a}nen}, {Page}, {Pedersen}, {Pozanenko},
  {Reeves}, {Rumyantsev}, {Shahbaz}, {Sharapov}, {Sollerman}, {Starling},
  {Tanvir}, {Torstensson}, \& {Wiersema}}]{watson2006}
{Watson} D. {et~al.}, 2006, APJ, 652, 1011

\bibitem[{{Waxman} \& {Draine}(2000)}]{waxman2000}
{Waxman} E., {Draine} B.~T., 2000, APJ, 537, 796

\bibitem[{{Wiersema}(2011)}]{wiersema2011}
{Wiersema} K., 2011, MNRAS, 414, 2793

\bibitem[{{Wolfe}, {Gawiser} \& {Prochaska}(2005){Wolfe}, {Gawiser}, \&
  {Prochaska}}]{wolfe2005}
{Wolfe} A.~M., {Gawiser} E., {Prochaska} J.~X., 2005, ARAA, 43, 861

\bibitem[{{Woosley}(1993)}]{woosley1993}
{Woosley} S.~E., 1993, APJ, 405, 273

\bibitem[{{Woosley} \& {Heger}(2006)}]{woosley2006}
{Woosley} S.~E., {Heger} A., 2006, ApJ, 637, 914

\end{thebibliography}

\section*{Affiliations}
\noindent $^{1}$Astronomical Institute $''$Anton Pannekoek$''$, University of Amsterdam, P.O. Box 94249, 1090 GE Amsterdam, The Netherlands.\\
$^{2}$Leiden Observatory, Leiden University, P.O. Box 9513, 2300 RA Leiden, The Netherlands.\\
$^{3}$University of Leicester, Department of Physics and Astronomy, University Road, Leicester LE1 7RH, United Kingdom\\
$^{4}$Centre for Astrophysics and Cosmology, Science Institute, University of Iceland, Dunhaga 5, IS-107 Reykjavik, Iceland\\
$^{5}$Max Planck Institute for Extraterrestrial Physics, 85748 Garching bei M\"unchen, Germany\\
$^{6}$Harvard-Smithsonian Center for Astrophysics, 60 Garden Street, Cambridge, MA 02138, USA\\
$^{7}$INAF - Osservatorio Astronomico di Brera, Via E. Bianchi 46, I-23807 Merate (LC), Italy\\
$^{8}$INAF-Osservatorio Astronomico di Roma, Via Frascati 33, I-00040 Monteporzio Catone, Italy \\
$^{9}$ASI-Science Data Centre, Via Galileo Galilei, I-00044 Frascati, Italy\\
$^{10}$Laboratoire GEPI, Observatoire de Paris, CNRS-UMR8111, Univ Paris Diderot, 5 place Jules Janssen, 92195 Meudon France\\
$^{11}$Dark Cosmology Centre, Niels Bohr Institute, University of Copenhagen, Juliane Maries Vej 30, DK-2100 Copenhagen, Denmark\\
$^{12}$APC, Astroparticules et Cosmologie, Universite Paris Diderot, CNRS/IN2P3, CEA/Irfu, Observatoire de Paris, Sorbonne Paris Cite, 10, rue Alice Domon et Leonie Duquet, 75205 Paris Cedex 13, France\\
$^{13}$Faculty of Mathematics and Physics, University of Ljubljana, Jadranska 19, SI-1000 Ljubljana, Slovenia\\
$^{14}$Centre of Excellence SPACE-SI, A\v{s}ker\v{c}eva cesta 12, SI-1000 Ljubljana, Slovenia\\
$^{15}$Astrophysics Research Institute, Liverpool John Moores University, Twelve Quays House, Egerton Wharf, Birkenhead, CH41 1LD, United Kingdom\\
$^{16}$Space Research Institute (IKI), 8432 Profsoyuznaya str., Moscow, 117997, Russia\\
$^{17}$Instituto de Astrof\'isica de Andaluc\'ia (IAA-CSIC), Glorieta de la Astronom\'ia s/n, E-18008, Granada, Spain

\appendix
\section{Tables with line transitions}

\begin{table*}
\flushleft

\caption{Lines from the ground state of Fe\,{\sc ii} (electron configurations $3d^6(^5D)4s - 3d^6(^5D)4p$) and its fine-structure levels, that could in principle be detected in the spectrum of \thisgrb, because the transition would lie within the GMOS spectral range and the oscillator strength $f_{\lambda}$ is not smaller than about $0.005$ (except for the $^6D_{9/2}$ resonance lines). The lines of which the wavelengths are in boldface have been used in the modelling reported in Section~\ref{sec:varyfinestruc}. $W_\lambda$ are measured equivalent widths per epoch, converted to the rest frame. }
\label{tab:Fetrans}
\begin{tabular}{l@{\hspace{2mm}}l@{\hspace{2mm}}l@{\hspace{2mm}}r@{\hspace{2mm}}l@{\hspace{2mm}}l@{\hspace{2mm}}l@{\hspace{2mm}}l@{\hspace{2mm}}l@{\hspace{2mm}}l} 
\hline

lower level		&$\lambda_{\textrm{vac}}$ (\AA)&$f_{\lambda}$ &$\log gf$		& $J_i -J_k$	& terms				& $W_{\lambda}(\mathrm{epoch1})$ & $W_{\lambda}(\mathrm{epoch2})$ & $W_{\lambda}(\mathrm{xsh})$ & $W_{\lambda}(\mathrm{epoch3})$ \\
\hline
$^6D_{9/2}$	&1608.4509     			&0.05399   & -0.268		& $9/2 - 7/2$	& $a^6D -y^6P^{\circ}$	& \multicolumn{4}{l}{possibly present, but blended with Al\,{\sc ii}$\,\lambda1670$ at $z=1.318$. }\\	
0 cm$^{-1}$	&1611.2004     			&0.00136   &-1.867		& $9/2 - 7/2$	& $a^6D -z^4F^{\circ}$	& \multicolumn{4}{l}{not detected.}\\	
			&\bf{2249.8754}     		&0.00219    &-1.660		& $9/2 - 7/2$	& $a^6D -z^4D^{\circ}$	& $0.129 \pm 0.034$ & $0.151 \pm 0.040$ & $0.143 \pm 0.044$ & $0.093 \pm 0.138$\\	
			&\bf{2260.7793}     		&0.00262    &-1.582		& $9/2 - 9/2$	& $a^6D -z^4F^{\circ}$	& $0.152 \pm 0.034$ & $0.168 \pm 0.040$ & $0.074 \pm 0.060$ & $0.256 \pm 0.125$\\	
			&2344.2129				&0.12523 &    0.098		& $9/2 - 7/2$	& $a^6D -z^6P^{\circ}$	& $0.748 \pm 0.047^a$ & $0.798 \pm 0.054^a$ & $0.704 \pm 0.189^a$ & $0.557 \pm 0.134^a$ \\
			&2374.4604     			&0.03297   &-0.482		& $9/2 - 9/2$	& $a^6D -z^6F^{\circ}$	& $0.590 \pm 0.045$ & $^b$ & $^b$ & $^b$\\		
			&2382.7641     			&0.34321     &0.536		& $9/2 - 11/2$	& $a^6D -z^6F^{\circ}$	& $0.683 \pm 0.045$ & $^b$ & $^{b,\,c}$ & $^b$\\		
			&\bf{2586.6495}     		&0.07094   &-0.149		& $9/2 - 7/2$	& $a^6D -z^6D^{\circ}$	& $0.705 \pm 0.043$ & $0.727 \pm 0.048$ & $0.696 \pm 0.145$ & $0.629 \pm 0.107$\\		
			&2600.1722     			&0.24226    &0.384		& $9/2 - 11/2$	& $a^6D -z^6D^{\circ}$	& $0.780 \pm 0.045$ & $^b$ & $^{b,\,c}$ & $^b$\\	
\hline
\multicolumn{10}{l}{$^a$Blended with lines from $^6D_{1/2}$ and $^4F_{7/2}$, total $W_{\lambda}$ is reported.}\\
\multicolumn{10}{l}{$^b$$W_{\lambda}$ of this resonance line is kept fixed to the value obtained at the first epoch to see the possible variation of the (half-) blended fine-structure lines.}\\
\multicolumn{10}{l}{$^c$The line (blend) is clearly saturated, and cannot be fit with gaussian (components) in the X-shooter spectrum.}\\				

\hline
$^6D_{7/2}$	&\bf{2333.5147}   		&0.07776   &-0.206		& $7/2 - 5/2$	& $a^6D -z^6P^{\circ}$	&$0.168 \pm 0.034$ & $0.137 \pm 0.038$ & $0.051 \pm 0.189$ & $-0.134 \pm 0.126$\\	
384.790 cm$^{-1}$&\bf{2365.5508}     	&0.05285   &-0.374		& $7/2 - 7/2$	& $a^6D -z^6P^{\circ}$	& $0.185 \pm 0.036$ & $0.130 \pm 0.039$ & $0.008 \pm 0.222$ & $0.056 \pm 0.120$\\		
			&2383.7873     		&0.00557   &-1.351		& $7/2 - 5/2$	& $a^6D -z^6F^{\circ}$	& $0.087 \pm 0.031$ & $0.071\pm 0.034$ & $^{c}$ 			& $0.054 \pm 0.107$\\		
			&\bf{2389.3569}     		&0.08987    &-0.143		& $7/2 - 7/2$	& $a^6D -z^6F^{\circ}$	& $0.187 \pm 0.033$ & $0.216 \pm 0.040$ & $-0.085 \pm 0.303$ & $-0.056 \pm 0.112$\\		
			&\bf{2396.3551}     		&0.30779    &0.391		& $7/2 - 9/2$	& $a^6D -z^6F^{\circ}$	& $0.383 \pm 0.040^d$ & $0.348 \pm 0.043^d$ & $0.170 \pm 0.275^d$ & $0.314 \pm 0.114^d$\\	
			&2599.1457     		&0.10863    &-0.061		& $7/2 - 5/2$	& $a^6D -z^DF^{\circ}$	& $0.190 \pm 0.031$ & $0.096 \pm 0.030$ & $ ^{c} $ 				& $-0.019 \pm 0.096$ \\ 
			&\bf{2612.6536}     		&0.12485    &-0.001		& $7/2 - 7/2$	& $a^6D -z^6D^{\circ}$	& $0.297 \pm 0.035$ & $0.241 \pm 0.037$ & $0.012 \pm 0.106$ & $0.049 \pm 0.096$ \\		
			&\bf{2626.4503}     		&0.04266    &-0.467		& $7/2 - 9/2$	& $a^6D -z^6D^{\circ}$	& $0.144 \pm 0.030$ & $0.120 \pm 0.031$ & $0.095 \pm 0.118$ & $0.012 \pm 0.092$\\	
\hline
\multicolumn{10}{l}{$^d$Blended with $^6D_{5/2}\lambda 2396.1487$, but we assume that the contribution from this line to the total $W_{\lambda}$ is negligible.}\\				
\hline
$^6D_{5/2}$	&\bf{2328.1100}     		&0.03760    &-0.647		& $5/2 - 3/2$	& $a^6D -z^6P^{\circ}$	&$0.052 \pm 0.031$ & $0.031 \pm 0.036$ & $0.064 \pm 0.235$ & $0.082 \pm 0.132$\\	
667.683 cm$^{-1}$&2349.0215     		&0.08935    &-0.271		& $5/2 - 5/2$	& $a^6D -z^6P^{\circ}$	&$0.204 \pm 0.035^e$ & $0.196 \pm 0.038^e$ & $0.081 \pm 0.154^e$ & $-0.139\pm 0.114^e$\\		
			&2381.4877     		&0.03798    &-0.642		& $5/2 - 7/2$	& $a^6D -z^6P^{\circ}$	& $0.022 \pm 0.027$ & $-0.009 \pm 0.033$ & $^c$ & $-0.004 \pm 0.111$\\
			&2396.1487     		&0.01636    &-1.008		& $5/2 - 3/2$	& $a^6D -z^6F^{\circ}$	& \multicolumn{4}{l}{blend with $^6D_{7/2}\lambda2396.3551$, which dominates.}\\	
			&\bf{2399.9718}    		 &0.12522    &-0.124		& $5/2 - 5/2$	& $a^6D -z^6F^{\circ}$	& $0.169 \pm 0.032$ & $0.163 \pm 0.037$ & $0.071 \pm 0.255$ & $0.221 \pm 0.124$\\
			&2405.6173     		&0.25103    &0.178		& $5/2 - 7/2$	& $a^6D -z^6F^{\circ}$	&$0.254 \pm 0.036^f$&$0.203 \pm 0.039^f$&$0.005 \pm 0.205^f$&$-0.076 \pm 0.102^f$\\		
			&2607.8658     		&0.12304    &-0.132		& $5/2 - 3/2$	& $a^6D -z^6D^{\circ}$	&$0.308 \pm 0.040^g$ & $0.242 \pm 0.043$ & $0.083 \pm 0.108$ & $0.007 \pm 0.131$\\	
			&\bf{2618.3984}     		&0.04800    &-0.541 		& $5/2 - 5/2$	& $a^6D -z^6P^{\circ}$	&$0.112 \pm 0.029$ & $0.057 \pm 0.031$ & $0.127 \pm 0.109$ & $-0.038 \pm 0.114$\\	
			&2632.1077     		&0.08240    &-0.306		& $5/2 - 7/2$	& $a^6D -z^6D^{\circ}$	&$0.252 \pm 0.033^h$ & $0.169 \pm 0.035^g$ & $-0.015 \pm 0.114^g$ & $0.086 \pm 0.088^g$ \\	

\hline
\multicolumn{10}{l}{$^e$Blended with $^4F_{9/2}\lambda2348.8346$, this is the total $W_{\lambda}$.}\\	
\multicolumn{10}{l}{$^f$Blended with $^6D_{3/2}\lambda2405.1626$, this is the total $W_{\lambda}$.}\\	
\multicolumn{10}{l}{$^g$Blended with Mn{\,\sc ii} $\lambda2606$, but this line's $W_{\lambda}$ is fixed after the first epoch.}\\
\multicolumn{10}{l}{$^h$Blended with $^6D_{3/2}\lambda2631.8314$, this is the total $W_{\lambda}$.}\\				
			
\hline
$^6D_{3/2}$	&\bf{2338.7238}     			&0.09840    &-0.405	& $3/2 - 3/2$	& $a^6D -z^6P^{\circ}$	&$0.090 \pm 0.032$ & $0.057 \pm 0.036$ & $0.127 \pm 0.174$ & $0.062 \pm 0.128$\\	
862.613 cm$^{-1}$&2359.8270    	 	&0.06788    &-0.566	& $3/2 - 5/2$	& $a^6D -z^6P^{\circ}$	&\multicolumn{4}{l}{part of a strongly blended complex, with a contribution of an unidentified line.}\\		
			&2405.1626     			&0.02784    &-0.953	& $3/2 - 1/2$	& $a^6D -z^6F^{\circ}$	&\multicolumn{4}{l}{blend with $^6D_{5/2}\lambda2405.6173$, which dominates.}\\		
			&\bf{2407.3934}     			&0.15988    &-0.194	& $3/2 - 3/2$	& $a^6D -z^6F^{\circ}$	&$0.122 \pm 0.032$ & $0.078 \pm 0.033$ & $0.014 \pm 0.199$ & $0.014 \pm 0.112$\\		
			&2411.2522     			&0.21967    &-0.056	& $3/2 - 5/2$	& $a^6D -z^6F^{\circ}$	& $0.227 \pm 0.035^i$&$0.115\pm 0.037^i$& $0.181 \pm 0.170^i$& $0.031 \pm 0.089^i$\\		
			&\bf{2614.6047}     			&0.11018    &-0.356	& $3/2 - 1/2$	& $a^6D -z^6D^{\circ}$	&$0.119 \pm 0.029$ & $0.091 \pm 0.032$ & $0.007 \pm 0.105$ & $0.055 \pm 0.095$\\		
			&2621.1906     			&0.00364    &-1.837 	& $3/2 - 3/2$	& $a^6D -z^6D^{\circ}$	& $0.047 \pm 0.028$& $-0.016 \pm 0.032$& $0.035 \pm 0.114$&$0.022 \pm0.084$\\	
			&2631.8314     			&0.12430    &-0.303	& $3/2 - 5/2$	& $a^6D -z^6D^{\circ}$	& \multicolumn{4}{l}{blended with $^6D_{5/2}\lambda2632.1077$, see there for the total $W_{\lambda}$.}\\	
\hline
\multicolumn{10}{l}{$^i$Blended with $^6D_{1/2}\lambda2411.8009$, this is the total $W_{\lambda}$.}\\	
\hline
$^6D_{1/2}$	&2344.9996     			&0.16819    &-0.473	& $1/2 - 3/2$	& $a^6D -z^6P^{\circ}$	& \multicolumn{4}{l}{blended with $^6D_{9/2}\lambda2344.214$ and $^4F_{7/2}\lambda2344.6789$, see there for the total $W_{\lambda}$.}\\	
977.053 cm$^{-1}$&2411.8009     	&0.22500    &-0.347	& $1/2 - 1/2$	& $a^6D -z^6F^{\circ}$	& \multicolumn{4}{l}{blended with $^6D_{3/2}\lambda2411.2522$, see there for the total $W_{\lambda}$.}\\		
			&\bf{2414.0437}     			&0.19047    &-0.419	& $1/2 - 3/2$	& $a^6D -z^6F^{\circ}$	&$0.060 \pm 0.030$ & $0.028 \pm 0.034$ & $-0.081 \pm 0.159$ & $0.073 \pm 0.125$\\		
			&\bf{2622.4513}     		&0.05454    &-0.962	& $1/2 - 1/2$	& $a^6D -z^6D^{\circ}$	& $0.036 \pm 0.025$ & $0.024 \pm 0.029$ & $0.114 \pm 0.109$ & $0.116 \pm 0.114$ \\		
			&\bf{2629.0769}     		&0.17451    &-0.457	& $1/2 - 3/2$	& $a^6D -z^6D^{\circ}$	& $0.079 \pm 0.027$ & $0.033 \pm 0.031$ & $0.008 \pm 0.172$ & $0.119 \pm 0.109$\\		

\hline
\end{tabular}
\end{table*}

\begin{table*}
\flushleft

\caption{The same as Table~\ref{tab:Fetrans}, only for the lines from the metastable states of Fe\,{\sc ii} with electron configuration $3p^63d^7$.}
\label{tab:Fetrans2}
\begin{tabular}{l@{\hspace{2mm}}l@{\hspace{2mm}}l@{\hspace{2mm}}r@{\hspace{2mm}}l@{\hspace{2mm}}l@{\hspace{2mm}}l@{\hspace{2mm}}l@{\hspace{2mm}}l@{\hspace{2mm}}l} 
\hline

lower level		&$\lambda_{\textrm{vac}}$ (\AA)&$f_{\lambda}$ &$\log gf$	& $J_i -J_k$	& terms			& $W_{\lambda}(\mathrm{epoch1})$ & $W_{\lambda}(\mathrm{epoch2})$ & $W_{\lambda}(\mathrm{xsh})$ & $W_{\lambda}(\mathrm{epoch3})$ \\
\hline

$^4F_{9/2}$	&1610.9234     &0.00914    &-1.039		& $9/2 - 9/2$	& $a^4F -y^4G^{\circ}$	&\multicolumn{4}{l}{not detected, hardly in range, noisy region.} \\	
1872.567 cm$^{-1}$&1637.3998     &0.01418    &-0.848		& $9/2 - 7/2$	& $a^4F -x^4D^{\circ}$	&\multicolumn{4}{l}{not detected, noisy region with many possible lines.} \\		
			&1702.0453     &0.05316    &-0.274		& $9/2 - 11/2$	& $a^4F -z^4G^{\circ}$	&\multicolumn{4}{l}{possibly detected, but shifted and blended with Ni{\,\sc ii}$\lambda1703$, noisy region.} \\		
			&2332.0235  &0.01787    &-0.748		& $9/2 - 7/2$	& $a^4F -z^4F^{\circ}$	& $0.052 \pm 0.029$ & $0.085 \pm 0.037$ & $0.109 \pm 0.199$ & $0.103 \pm 0.123$\\		
			&2348.8346     &0.03653    &-0.437		& $9/2 - 7/2$	& $a^4F -z^4D^{\circ}$	&\multicolumn{4}{l}{blended with  $^6D_{5/2}\lambda 2349.0215$, see there for total $W_{\lambda}$.}\\	
			&2360.7213     &0.02289    &-0.640		& $9/2 - 9/2$	& $a^4F -z^4F^{\circ}$	& \multicolumn{4}{l}{part of a strongly blended complex, with a contribution of an unidentified line.}\\		
\hline		
$^4F_{7/2}$	&1625.5231	&0.02481	&-0.702		& $7/2 - 9/2$	& $a^4F -y^4G^{\circ}$	&\multicolumn{4}{l}{not detected, noisy region.} \\	 
2430.097 cm$^{-1}$&	2344.6790&0.01669 &   -0.874		& $7/2 - 5/2$	& $a^4F -z^4F^{\circ}$	&\multicolumn{4}{l}{blended with $^6D_{9/2}\lambda2344.214$ and $^6D_{1/2}\lambda2344.9996$, see there for the total $W_{\lambda}$.} \\	
			&2361.0161     &0.03053  &  -0.612		& $7/2 - 5/2$	& $a^4F -z^4D^{\circ}$	&\multicolumn{4}{l}{part of a strongly blended complex, with a contribution of an unidentified line.}\\		
			&2380.0023     &0.01732   & -0.858		& $7/2 - 7/2$	& $a^4F -z^4D^{\circ}$	&\multicolumn{4}{l}{not detected.} \\	
\hline	
$^4F_{5/2}$	&1633.9093     &0.02620   & -0.803		& $5/2 - 7/2$	& $a^4F -y^4G^{\circ}$	&\multicolumn{4}{l}{not detected, noisy region with many possible lines.} \\	
2837.950 cm$^{-1}$&2369.3195     &0.02918    &-0.757		& $5/2 - 3/2$	& $a^4F -z^4D^{\circ}$	&\multicolumn{4}{l}{not detected.} \\	
		 	&2383.9718     &0.02215   & -0.876		& $5/2 - 5/2$	& $a^4F -z^4D^{\circ}$	&\multicolumn{4}{l}{blended with many lines including the saturated $^6D_{9/2}\lambda2382.7641$.} \\	
\hline	
$^4F_{3/2}$	&2375.9188     &0.03512    &-0.852	& $3/2 - 1/2$	& $a^4F -y^4D^{\circ}$	& $0.061 \pm 0.034^j$ & $0.066 \pm0.034^j$ & $-0.147 \pm 0.202^j$ & $-0.079 \pm 0.100^j$ \\	
3117.461 cm$^{-1}$&2385.1148     &0.02311    &-1.034		& $3/2 - 3/2$	& $a^4F -y^4D^{\circ}$	&\multicolumn{4}{l}{not detected, close to complex blend.} \\	
\hline	
\multicolumn{10}{l}{$^j$Half-blended with $^6D_{9/2}\lambda2374.4604$, which's $W_{\lambda}$ is fixed from the first epoch.}\\	
\hline
\end{tabular}
\end{table*}

\begin{table*}
\flushleft
\caption{The same as Table~\ref{tab:Fetrans}, only for the lines from the metastable states of Fe\,{\sc ii} with electron configuration $3d^6(^5D)4s$.}
\label{tab:Fetrans_metastab}
\begin{tabular}{l@{\hspace{2mm}}l@{\hspace{2mm}}l@{\hspace{2mm}}r@{\hspace{2mm}}l@{\hspace{2mm}}l@{\hspace{2mm}}l@{\hspace{2mm}}l@{\hspace{2mm}}l@{\hspace{2mm}}l} 
\hline

lower level		&$\lambda_{\textrm{vac}}$ (\AA)&$f_{\lambda}$ &$\log gf$	& $J_i -J_k$	& terms			& $W_{\lambda}(\mathrm{epoch1})$ & $W_{\lambda}(\mathrm{epoch2})$ & $W_{\lambda}(\mathrm{xsh})$ & $W_{\lambda}(\mathrm{epoch3})$ \\

\hline

$^4D_{7/2}$	&1635.4005     &0.06857    &-0.261		& $7/2 - 5/2$	& $a^4D -x^4P^{\circ}$	&\multicolumn{4}{l}{heavily blended, noisy region.} \\		
7955.299 cm$^{-1}$&\bf{2563.3039}     &0.13520     &0.034		& $7/2 - 5/2$	& $a^4D -z^4P^{\circ}$	&$0.037 \pm 0.027$ & $0.064 \pm 0.032$ & $-0.052 \pm 0.248$ & $0.047 \pm 0.110$
\\				
			&2715.2173     &0.04543    &-0.440		& $7/2 - 5/2$	& $a^4D -z^4D^{\circ}$	&\multicolumn{4}{l}{not detected.} \\		
			&\bf{2740.3577}     &0.25107     &0.303		& $7/2 - 7/2$	& $a^4D -z^4D^{\circ}$	&$0.056 \pm 0.026$ & $0.027 \pm 0.029$ & $0.018 \pm 0.099$ & $0.114 \pm 0.094$\\		
			&\bf{2756.5512}   &0.30759     &0.391		& $7/2 - 9/2$	& $a^4D -z^4F^{\circ}$	&$0.050 \pm 0.025$ & $0.056 \pm 0.030$ & $0.097 \pm 0.114$ & $-0.045\pm 0.104$\\	
\hline	
$^4D_{5/2}$	&1641.7631     &0.04742    &-0.546		& $5/2 - 3/2$	& $a^4D -x^4P^{\circ}$	& \multicolumn{4}{l}{blended, noisy region.}\\		
8391.938 cm$^{-1}$&1647.1625   &  0.02026  &  -0.915		& $5/2 - 5/2$	& $a^4D -x^4P^{\circ}$	& \multicolumn{4}{l}{blended, noisy region.}\\	
			& 2564.2441     &0.09332    &-0.252	& $5/2 - 3/2$	& $a^4D -z^4P^{\circ}$	&\multicolumn{4}{l}{not detected.} \\		
			&2592.3179     &0.05219    &-0.504		& $5/2 - 5/2$	& $a^4D -z^4P^{\circ}$	&\multicolumn{4}{l}{not detected.}\\		
			& 2728.3467     &0.06890    &-0.384		& $5/2 - 3/2$	& $a^4D -z^4D^{\circ}$	&\multicolumn{4}{l}{not detected.} \\		
			&2747.7942     &0.18791     &0.052		& $5/2 - 5/2$	& $a^4D -z^4D^{\circ}$	&\multicolumn{4}{l}{part of strongly blended complex.} \\		
			& 2750.1347     &0.32808   &  0.294		& $5/2 - 5/2$	& $a^4D -z^4F^{\circ}$	&\multicolumn{4}{l}{blended with $^4D_{3/2}\lambda 2749.9938$ and $^4D_{1/2}\lambda 2750.2989$.}\\		
\hline	
$^4D_{3/2}$	& 2567.6820   &  0.05684   & -0.643		& $3/2 - 1/2$	& $a^4D -z^4P^{\circ}$	&\multicolumn{4}{l}{not detected.}\\		
8680.454 cm$^{-1}$& 2583.3564   &  0.08345 &   -0.477		& $3/2 - 3/2$	& $a^4D -z^4P^{\circ}$	&\multicolumn{4}{l}{not detected.}\\	
			&2731.5431     &0.02964    &-0.926		& $3/2 - 3/2$	& $a^4D -z^4F^{\circ}$	&\multicolumn{4}{l}{not detected.}\\		
			&2737.7761     &0.06574    &-0.580		& $3/2 - 1/2$	& $a^4D -z^4D^{\circ}$	&\multicolumn{4}{l}{not detected.}\\		
			& 2747.2958     &0.35475    & 0.152	& $3/2 - 5/2$	& $a^4D -z^4F^{\circ}$	&\multicolumn{4}{l}{part of strongly blended complex.} \\		
			&2749.9938     &0.13492    &-0.268	& $3/2 - 3/2$	& $a^4D -z^4D^{\circ}$	&\multicolumn{4}{l}{blended with $^4D_{5/2}\lambda 2750.1344$ and $^4D_{1/2}\lambda 2750.2988$.} \\
	
\hline	
$^4D_{1/2}$	&2578.6940     &0.12362   & -0.607		& $1/2 - 1/2$	& $a^4D -z^4P^{\circ}$	&\multicolumn{4}{l}{not detected.} \\		
8846.768 cm$^{-1}$&2744.0089   &  0.44026 &   -0.055		& $1/2 - 3/2$	& $a^4D -z^4F^{\circ}$	&\multicolumn{4}{l}{possibly detected, but half blended Mg{\,\sc i }$\lambda 2853$ at $z=1.315$.}\\	
			&  2750.2989   &  0.12021  &  -0.619	& $1/2 - 1/2$	& $a^4D -z^4D^{\circ}$	&\multicolumn{4}{l}{blended with $^4D_{5/2}\lambda 2750.1347$ and $^4D_{3/2}\lambda 2749.9938$.} \\		
\hline	

\end{tabular}
\end{table*}

\begin{table*}
\flushleft

\caption{The same as Table~\ref{tab:Fetrans}, only for the lines from the metastable states of Ni\,{\sc ii} with electron configurations $3p^63d^8(^3F)4s$.}
\label{tab:Nitrans_metastab}
\begin{tabular}{l@{\hspace{2mm}}l@{\hspace{2mm}}l@{\hspace{2mm}}r@{\hspace{2mm}}l@{\hspace{2mm}}l@{\hspace{2mm}}l@{\hspace{2mm}}l@{\hspace{2mm}}l@{\hspace{2mm}}l} 
\hline

lower level		&$\lambda_{\textrm{vac}}$ (\AA)&$f_{\lambda}$ &$\log gf$	& $J_i -J_k$	& terms			& $W_{\lambda}(\mathrm{epoch1})$ & $W_{\lambda}(\mathrm{epoch2})$ & $W_{\lambda}(\mathrm{xsh})$ & $W_{\lambda}(\mathrm{epoch3})$ \\

\hline

$^4F_{9/2}$	&\bf{2166.2300}		& 0.09990	&0.000		& $9/2 - 9/2$	& $^4F -^2F^{\circ}$	&$0.070 \pm 0.035$ & $0.191 \pm 0.045$ & $0.197 \pm 0.046$ & $0.026 \pm 0.145$\\		
8393.90 cm$^{-1}$&\bf{2217.1676}	&0.17865	&0.252		& $9/2 - 11/2$	& $^4F -^4G^{\circ}$	&$0.110 \pm 0.035$ & $0.222 \pm 0.042$ & $0.150 \pm 0.043$ & $0.227 \pm 0.143$\\
			&\bf{2223.6419}		&0.04789	&-0.320		& $9/2 - 9/2$	& $^4F -^4G^{\circ}$	&$0.019 \pm 0.031$ & $0.116 \pm 0.039$ & $0.042 \pm 0.046$ & $-0.060 \pm 0.136$\\		
			&2316.7481		&0.11073	&0.044		& $9/2 - 7/2$	& $^4F -^4D^{\circ}$	&\multicolumn{4}{l}{blended with a skyline.} \\	
\hline	

\end{tabular}
\end{table*}

\end{document}